\documentclass{aa}  
\usepackage{natbib}
\usepackage{tikz}
\usepackage{booktabs}
\usepackage{graphicx,txfonts}
\usepackage{orcidlink}
\usepackage{longtable}
\usepackage{subcaption}

\usetikzlibrary{svg.path}
\usetikzlibrary{shapes,arrows,positioning}
\let\orcidicon\orcidlink

\usepackage{hyperref} 
\hypersetup{
    colorlinks=true,
    citecolor=blue,
    linkcolor=blue,
    urlcolor=blue,
    }

\definecolor{color1bg}{HTML}{D7191C}
\definecolor{color2bg}{HTML}{FDAE61}
\definecolor{color3bg}{HTML}{FFFFBF}
\definecolor{color4bg}{HTML}{ABD9E9}
\definecolor{color5bg}{HTML}{2C7BB6}
    
\definecolor{cadmiumred}{rgb}{0.89, 0.0, .138}
\newcommand{\red}[1]{\textcolor{cadmiumred}{\bf{#1}}}

\newcounter{lenote}

\graphicspath{ {./images/} }
\makeatletter
\renewcommand*\aa@pageof{, page \thepage{} of \pageref*{LastPage}}
\makeatother

\begin{document} 

\authorrunning{Cavuoti, De Cicco et al.}
\titlerunning{Identification of problematic epochs in Astronomical Time Series}

   \title{Identification of problematic epochs in astronomical time series through transfer learning}


   \author{Stefano Cavuoti
          \inst{1,2,\orcidicon{0000-0002-3787-4196}}\fnmsep\thanks{The first two authors share the first authorship and are in alphabetical order. \\e-mail addresses:\\  \href{mailto:stefano.cavuoti@inaf.it}{stefano.cavuoti@inaf.it} and \href{mailto:demetra.decicco@unina.it}{demetra.decicco@unina.it}}
          \and
          Demetra De Cicco\inst{3,1,4,\orcidicon{0000-0001-7208-5101}}\fnmsep$^\star$
          \and
          Lars Doorenbos\inst{5,\orcidicon{0000-0002-0231-9950}}
          \and\\ 
          Massimo Brescia\inst{3,1,2,\orcidicon{0000-0001-9506-5680}}
          \and 
          Olena Torbaniuk\inst{6,7,\orcidicon{0000-0003-4465-2564}}
          \and
          Giuseppe Longo\inst{3,\orcidicon{0000-0002-9182-8414}}
          \and
          Maurizio Paolillo\inst{3,1,2,\orcidicon{0000-0003-4210-7693}}
          }

   \institute{INAF - Osservatorio Astronomico di Capodimonte, via Moiariello 16, 80131 Napoli, Italy 
         \and
         INFN -- Section of Naples, via Cinthia 9, 80126 Napoli, Italy 
         \and
             Department of Physics, University of Napoli ``Federico II'', via Cinthia 9, 80126 Napoli, Italy
        \and 
        Millennium Institute of Astrophysics (MAS), Nuncio Monse\~nor Sotero Sanz 100, Providencia, Santiago, Chile
        \and
        AIMI, ARTORG Center, University of Bern, Murtenstrasse 50, Bern, CH-3008, Switzerland
        \and
        Department of Physics and Astronomy ‘Augusto Righi’, University of Bologna, via Piero Gobetti 93/2, I-40129, Bologna, Italy;
        \and
        INAF — Osservatorio di Astrofisica e Scienza dello Spazio di Bologna, via Piero Gobetti 101, I-40129, Bologna, Italy;
             }

   \date{Received March 28, 2024; accepted May 06, 2024}

 
  \abstract
   {}
   {We present a novel method for detecting outliers in astronomical time series based on the combination of a deep neural network and a k-nearest neighbor algorithm with the aim of identifying and removing problematic epochs in the light curves of astronomical objects.
   }
   {We used an EfficientNet network pretrained on ImageNet as a feature extractor and performed a k-nearest neighbor search in the resulting feature space to measure the distance from the first neighbor for each image. If the distance was above the one obtained for a stacked image, we flagged the image as a potential outlier. }
   {We applied our method to a time series obtained from the VLT Survey Telescope monitoring campaign of the Deep Drilling Fields of the Vera C. Rubin Legacy Survey of Space and Time \thanks{Observations were provided by the ESO programs 088.D-4013, 092.D-0370, and 094.D-0417 (PI G. Pignata).}. We show that our method can effectively identify and remove artifacts from the VST time series and improve the quality and reliability of the data. This approach may prove very useful in light of the amount of data that will be provided by the LSST, which will prevent the inspection of individual light curves. We also discuss the advantages and limitations of our method and suggest possible directions for future work.
   \\The code is available \href[pdfnewwindow=true]{https://github.com/cavuoti/AnomalyInTimeSeries}{here}.
   }
   {}

   \keywords{Machine learning -- Galaxies: active  -- Techniques: image processing -- Methods: data analysis -- Artificial Intelligence 
               }

   \maketitle
%
\section{Introduction}
\label{introduction}

Astronomical time series are sequences of data that represent the variation of a physical quantity over time in the observation of celestial objects or phenomena. They are essential for studying and discovering the properties and dynamics of various astrophysical sources, such as stars, planets, supernovae, black holes, and galaxies~\citep{Scargle97,Aigrain23}.

However, an astronomical time series analysis poses several challenges, such as the presence of noise, gaps, outliers, and artifacts in the data. Artifacts are unwanted signals that do not reflect the true nature of the observed object and are caused by external factors, such as instrumental errors, atmospheric conditions, peculiar events (e.g., cosmic rays, satellite tracks), and contaminant sources ~\citep{Li18,Malz19}. Artifacts can severely affect the quality and reliability of data and may lead to false or misleading conclusions. Therefore, it is crucial to detect and remove artifacts from astronomical time series before performing any further analysis.

One specific case of astronomical time series are those obtained from the VLT Survey Telescope (VST), a 2.6-meter optical telescope located at the Paranal Observatory in Chile. The VST is designed to perform wide-field imaging surveys of the southern sky, covering a field of view (FoV) of one square degree with a pixel scale of 0.214\arcsec~\citep{CapaccioliSchipani15}. The VST time series are useful for studying various topics, such as variable stars, transient events, Solar System objects, and cosmology~\citep{Cappellaro15,Falocco15,decicco15,decicco19,decicco21,Botticella17,Fu18,Liu20,Poulain20}.

The VST time series, as all astronomical data, are affected by artifacts, such as bad pixels, saturation, cosmic rays, ghost images, and background fluctuations. These artifacts can hamper the detection and characterization of faint or fast-varying sources and decrease the scientific value of the data. Data reduction and calibration methods (see, e.g.,~\citealt{grado12}) are expected to correct and mask such artifacts or, when this proves difficult, flag them to allow easy filtering in the data analysis phase. However, it is not unusual to find residual defects in the reduced data, especially in time series, where we cannot use time averaging to remove many of the aforementioned problems. Therefore, it is important to develop improved methods for identifying and flagging these artifacts in order to limit their impact on scientific results.

One possible way to tackle this problem is to use transfer learning, a technique in machine learning that allows one to leverage the knowledge learned from a source domain to improve the performance on a target domain. Transfer learning can be useful when the target domain has limited or scarce data or when the source domain has some similarities or relevance to the target domain, as happens in astronomy (at least in terms of labels;~\citealt{Awang20,Martinazzo21}), malware classification~\citep{Prima20}, Earth science~\citep{Zou18}, and medicine~\citep{Ding19, Esteva17,Kim21, Menegola17}. 

In this work, we propose the use of an EfficientNet network trained on ImageNet as an anomaly detector for astronomical time series. An EfficientNet network~\citep{Tan19} is a convolutional neural network (CNN) architecture that scales up the network depth, width, and resolution in a balanced and efficient way using a compound coefficient. ImageNet is a large-scale dataset often used for such a purpose~\citep{deng09}, and it contains around 1.3 million images, where the original task was to classify each image into one of 1,000 classes.

The idea is to use the EfficientNet network as a feature extractor, as proposed in~\cite{Doorenbos22}, in the detection of active galactic nuclei (AGN) and to then use an algorithm for the detection of sources deviating from the standard behavior. 

This approach can improve and automatize the identification and removal of problematic epochs in the light curves without the need for visual inspection. This process can therefore improve the reliability of the light curves and the results obtained with their analysis. We expect this to be highly relevant in light of the Vera C. Rubin Legacy Survey of Space and Time (LSST; \citealt{ivezic2019}).

The paper is organized as follows: In Sec.~\ref{sec:data}, we describe the data used for this work, while in Sec.~\ref{sec:methdod}, we present the method. In Sec.~\ref{sec:results}, we discuss the results, and finally in Sec.~\ref{sec:conclusion} we draw our conclusions.

\section{Data}\label{sec:data}

As a benchmark, we used the same dataset as~\cite{decicco19,decicco21,decicco22}. The dataset consists of $r$-band observations of the COSMOS field performed with the VST spanning three observing seasons (hereafter, seasons) from December 2011 to March 2015 and are part of a long-term effort to monitor the LSST Deep Drilling Fields before the start of the Vera Rubin Telescope operations. The VST is a 2.6-meter optical telescope that covers a FoV of $1^\circ\times1^\circ$ with a single pointing (the pixel scale is $0.214$\arcsec/pixel). The three seasons include 54 visits in total and with two gaps; more details about the dataset are given in Table \ref{tab:dataset} (adapted from \href[pdfnewwindow=true]{https://www.aanda.org/articles/aa/full_html/2019/07/aa35659-19/T1.html}{Table 1} of~\cite{decicco19}). As explained in~\cite{decicco19}, they excluded 11 of the 65 visits that constitute the full dataset.

\begin{table*}[!htb]
\centering
\caption{VST-COSMOS dataset used in this work. The four columns report the visit number, OB identification number, observing date, and seeing FWHM, respectively, for the 54 visits used for the present analysis; visits are listed in chronological order. All the visits were obtained by combining five exposures, for a total exposure time of 1,800 s. However, visit 53 was obtained by combining ten exposures, for a total exposure time of 3,600 s. Adapted from \href[pdfnewwindow=true]{https://www.aanda.org/articles/aa/full_html/2019/07/aa35659-19/T1.html}{Table 1} of~\cite{decicco19}.}
\label{tab:dataset}      
\begin{minipage}{0.49\textwidth}
\centering
\begin{tabular}{c c c c}
\toprule
visit & OB-ID & obs. date & seeing (FWHM)\\
 & & & (arcsec)\\
\midrule
1 & $\mbox{611279}$ & 2011-Dec-18 & $0.64 $\\
2 & $\mbox{611283}$ & 2011-Dec-22 & $0.94 $\\
3 & $\mbox{611287}$ & 2011-Dec-27 & $1.04 $\\
4 & $\mbox{611291}$ & 2011-Dec-31 & $1.15 $\\
5 & $\mbox{611295}$ & 2012-Jan-02 & $0.67 $\\
6 & $\mbox{611299}$ & 2012-Jan-06 & $0.58 $\\ %
7 & $\mbox{611311}$ & 2012-Jan-18 & $0.62 $\\
8 & $\mbox{611315}$ & 2012-Jan-20 & $0.88 $\\   
9 & $\mbox{611319}$ & 2012-Jan-22 & $0.81 $\\   
10 & $\mbox{611323}$ & 2012-Jan-24 & $0.67 $\\
11 & $\mbox{611327}$ & 2012-Jan-27 & $0.98 $\\  
12 & $\mbox{611331}$ & 2012-Jan-29 & $0.86 $\\  
13 & $\mbox{611335}$ & 2012-Feb-02 & $0.86 $\\  
14 & $\mbox{611351}$ & 2012-Feb-16 & $0.50 $\\
15 & $\mbox{611355}$ & 2012-Feb-19 & $0.99 $\\  
16 & $\mbox{611359}$ & 2012-Feb-21 & $0.79 $\\  
17 & $\mbox{611363}$ & 2012-Feb-23 & $0.73 $\\  
18 & $\mbox{611367}$ & 2012-Feb-26 & $0.83 $\\  
19 & $\mbox{611371}$ & 2012-Feb-29 & $0.90 $\\  
20 & $\mbox{611375}$ & 2012-Mar-03 & $0.97 $\\  
21 & $\mbox{611387}$ & 2012-Mar-13 & $0.70 $\\  
22 & $\mbox{611391}$ & 2012-Mar-15 & $1.08 $\\  
23 & $\mbox{611395}$ & 2012-Mar-17 & $0.91 $\\  
24 & $\mbox{768813}$ & 2012-May-08 & $0.74 $\\  
25 & $\mbox{768817}$ & 2012-May-11 & $0.85 $\\  
26 & $\mbox{768820}$ & 2012-May-17 & $0.77 $\\  
27 & $\mbox{986611}$ & 2013-Dec-27 & $0.72 $\\  
    &                               &                       &             \\
\bottomrule
\end{tabular}
\end{minipage} \hfill
\begin{minipage}{0.49\textwidth}
\centering
\begin{tabular}{c c c c}
\toprule
visit & OB-ID & obs. date & seeing (FWHM)\\
 & & & (arcsec)\\
\midrule
28 & $\mbox{986614}$ & 2013-Dec-30 & $1.00 $\\  
29 & $\mbox{986617}$ & 2014-Jan-03 & $0.86 $\\  
30 & $\mbox{986620}$ & 2014-Jan-05 & $0.81 $\\  
31 & $\mbox{986626}$ & 2014-Jan-12 & $0.73 $\\  
32 & $\mbox{986630}$ & 2014-Jan-21 & $1.18 $\\
33 & $\mbox{986633}$ & 2014-Jan-24 & $0.80 $\\  
34 & $\mbox{986648}$ & 2014-Feb-09 & $1.28 $\\
35 & $\mbox{986652}$ & 2014-Feb-19 & $0.89 $\\  
36 & $\mbox{986655}$ & 2014-Feb-21 & $0.93 $\\  
37 & $\mbox{986658}$ & 2014-Feb-23 & $0.81 $\\  
38 & $\mbox{986661}$ & 2014-Feb-26 & $0.81 $\\  
39 & $\mbox{986664}$ & 2014-Feb-28 & $0.77 $\\  
40 & $\mbox{986670}$ & 2014-Mar-08 & $0.91 $\\  
41 & $\mbox{986674}$ & 2014-Mar-21 & $0.96 $\\  
42 & $\mbox{986677}$ & 2014-Mar-23 & $0.92 $\\  
43 & $\mbox{986680}$ & 2014-Mar-25 & $0.66 $\\
44 & $\mbox{1095777}$ & 2014-Mar-29 & $0.89 $\\ 
45 & $\mbox{1095783}$ & 2014-Apr-04 & $0.58 $\\ 
46 & $\mbox{986683}$ & 2014-Apr-07 & $0.61 $\\  
47 & $\mbox{1136410}$ & 2014-Dec-03 & $1.00 $\\ 
48 & $\mbox{1136457}$ & 2015-Jan-10 & $0.71 $\\ 
49 & $\mbox{1136481}$ & 2015-Jan-28 & $0.90 $\\ 
50 & $\mbox{1136490}$ & 2015-Jan-31 & $0.73 $\\ 
51 & $\mbox{1136503}$ & 2015-Feb-15 & $0.70 $\\ 
52 & $\mbox{1136531}$ & 2015-Mar-10 & $0.80 $\\ 
53 & $\mbox{1136540}$ & 2015-Mar-14 & $0.84 $\\ 
54 & $\mbox{1136543}$ & 2015-Mar-19 & $1.00 $\\ 
stacked & - & - & $0.67 $\\
\bottomrule
\end{tabular}
\end{minipage}
\end{table*}

The observations in the $r$-band were designed to have a three-day observing cadence, although the exact final cadence depended on the observational constraints. The single-visit depth is $r \lesssim 24.6$ mag for point sources, at a ${\sim}5\sigma$ confidence level. This makes our dataset particularly interesting for studies aimed at forecasting the performance of the LSST, as its single-visit depth is expected to be similar to our stacked COSMOS images. 
We refer the reader to~\cite{decicco15} for information on the reduction and combination of the exposures performed using the VST-Tube pipeline~\citep{grado12} as well as the source extraction and sample assembly. The VST-Tube magnitudes are in the AB system.

The sample contains 22,927 sources detected in at least 50\% of the visits in the dataset (i.e., they have at least 27 points in their light curves) with an average magnitude of $r \leq 23.5$ mag within a $1\arcsec$-radius aperture. We started with the same sample for our analysis, but while~\cite{decicco19} focused on the sub-sample of sources selected by either variability or multiwavelength properties, we used the entire sample, as done in~\cite{decicco21}. In fact, our aim is to efficiently identify the unreliable measurements in the full dataset before engaging in any classification effort that could be biased by photometric or aesthetic artifacts. These problematic sources are mainly satellite tracks, aesthetic defects, cosmic rays, saturation, and bad seeing together with objects blended with a neighbor in at least some of the visits (those with poor seeing), making it substantially harder to measure fluxes correctly. 
We decided to analyze and compare this sample, as it comes from the pipeline and dataset already presented in~\cite{decicco21}.

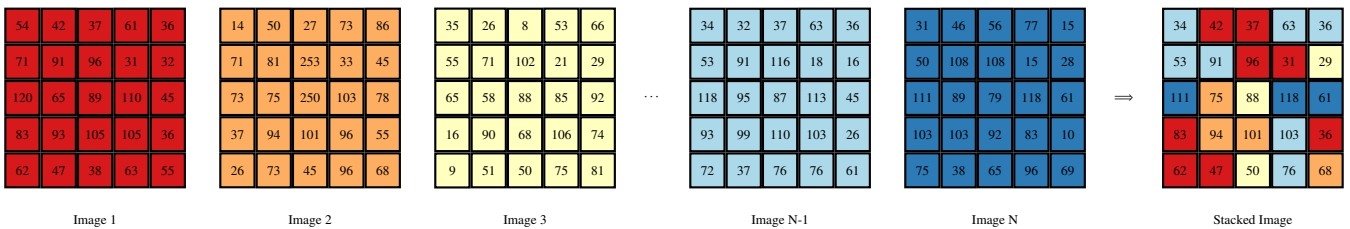
\begin{figure*}[!htb]
    \centering

\begin{tikzpicture}[thick,scale=0.49, every node/.style={scale=0.5}]

\usetikzlibrary{matrix,positioning}

\matrix (matrix1) [matrix of math nodes, nodes={draw=black, fill=color1bg!100, minimum size=9mm}]
{
  54 & 42 & 37 & 61 & 36 \\
  71 & 91 & 96 & 31 & 32 \\
  120 & 65 & 89 & 110 & 45 \\
  83 & 93 & 105 & 105 & 36 \\
  62 & 47 & 38 & 63 & 55 \\
};
\node[below=1ex of matrix1] {Image 1};

\matrix (matrix2) [matrix of math nodes, right=0.5em of matrix1, nodes={draw=black, fill=color2bg!100, minimum size=9mm}]
{
  14 & 50 & 27 & 73 & 86 \\
  71 & 81 & 253 & 33 & 45 \\
  73 & 75 & 250 & 103 & 78\\
  37 & 94 & 101 & 96 & 55\\
  26 & 73 & 45 & 96 & 68 \\
};
\node[below=1ex of matrix2] {Image 2};

\matrix (matrix3) [matrix of math nodes, right=0.5em of matrix2, nodes={draw=black, fill=color3bg!100, minimum size=9mm}]
{
  35 & 26 & 8 & 53 & 66 \\
  55 & 71 & 102 & 21 & 29 \\
  65 & 58 & 88 & 85 & 92 \\
  16 & 90 & 68 & 106 & 74 \\
  9 & 51 & 50 & 75 & 81 \\
};
\node[below=1ex of matrix3] {Image 3};

\node(puntini)[right=0.5em of matrix3] {$\cdots$};

\matrix (matrix4) [matrix of math nodes, right=0.5em of puntini, nodes={draw=black, fill=color4bg!100, minimum size=9mm}]
{
  34 & 32 & 37 & 63 & 36 \\
  53 & 91 & 116 & 18 & 16 \\
  118 & 95 & 87 & 113 & 45 \\
  93 & 99 & 110 & 103 & 26 \\
  72 & 37 & 76 & 76 & 61 \\
};
\node[below=1ex of matrix4] {Image N-1};

\matrix (matrix5) [matrix of math nodes, right=0.5em of matrix4, nodes={draw=black,  fill=color5bg!100, minimum size=9mm}]
{
  31 & 46 & 56 & 77 & 15 \\
  50 & 108 & 108 & 15 & 28 \\
  111 & 89 & 79 & 118 & 61 \\
  103 & 103 & 92 & 83 & 10 \\
  75 & 38 & 65 & 96 & 69 \\
};
\node[below=1ex of matrix5] {Image N};

\node(arrow)[right=0.5em of matrix5] {$\Longrightarrow$};

\matrix (medianMatrix) [matrix of math nodes, right=0.5em of arrow, nodes={draw=black,  minimum size=9mm}]
{
|[fill=color4bg!100]|  34 & |[fill=color1bg!100]| 42 & |[fill=color1bg!100]| 37 & |[fill=color4bg!100]| 63 & |[fill=color4bg!100]|36 \\
 |[fill=color4bg!100]| 53 & |[fill=color4bg!100]|91 & |[fill=color1bg!100]|96 & |[fill=color1bg!100]|31 &  |[fill=color3bg!100]|29 \\
   |[fill=color5bg!100]|111 &  |[fill=color2bg!100]|75 &  |[fill=color3bg!100]| 88 &  |[fill=color5bg!100]|118 &  |[fill=color5bg!100]| 61\\
  |[fill=color1bg!100]| 83 &  |[fill=color2bg!100]|94 &  |[fill=color2bg!100]| 101 &  |[fill=color4bg!100]|103 &  |[fill=color1bg!100]|36 \\
  |[fill=color1bg!100]| 62 &  |[fill=color1bg!100]|47 &  |[fill=color3bg!100]|50 &  |[fill=color4bg!100]|76 &  |[fill=color2bg!100]|68 \\
};
\node[below=1ex of medianMatrix] {Stacked Image};

\end{tikzpicture}

    \caption {Example derivation of our stacked image starting from individual images. The individual images are represented by simple 5x5 matrices. The median value for each cell is selected in order to derive the stacked image. This process is suitable to avoiding wrong values in single epochs. For example, the values above 250 in Image 2 have been removed completely, as when using the mean value, the presence of such values, although mitigated by the number of independent images, would affect the final result much more.}
    \label{fig:stacked}
\end{figure*}

As the reference for each object, we used the stacked image, that is, the median of all the exposures in the first season -- which covers a five-month baseline -- with a seeing full width at half maximum (FWHM) of less than $0.80\arcsec$. This choice ensured that the reference image is of higher quality than each individual epoch and that it does not include all the defects present in the whole dataset. A visual representation of the resulting image from single epochs is shown in Fig.~\ref{fig:stacked}. 

\section{Method}\label{sec:methdod}

We propose a novel method for detecting outliers in astronomical time series based on combining a deep neural network and a k-nearest neighbor (k-NN) algorithm. The preliminary phase is similar to what has been done in~\cite{Doorenbos22}. The main idea is to use an EfficientNet, a CNN~\citep{Tan19}, pretrained on ImageNet as a feature extractor and to perform a k-NN search in the resulting feature space to measure the distance to the first neighbor for each image. If the distance is above a certain threshold, we flag the image as a potential outlier.

Convolutional neural networks \citealt{LeCun90,lecun2015} are a class of deep neural networks designed for processing structured grid data, such as images. They have been proven to be highly effective in computer vision tasks, including image classification, object detection, and segmentation.
A typical CNN consists of multiple layers, including:

\begin{itemize}
  \item \textbf{Convolutional layers:} These layers apply convolution operations to their input, using learnable filters to extract features. Convolutional layers help capture hierarchical patterns in the input.

  \item \textbf{Pooling layers:} Pooling layers reduce the spatial dimensions of the input by downsampling. Common pooling operations include max pooling, which retains the maximum value in a region, and average pooling, which computes the average.

  \item \textbf{Fully connected layers:} These layers connect every neuron to every neuron in the previous and subsequent layers, and they help make predictions based on the extracted features.

  \item \textbf{Activation functions:} Non-linear activation functions, such as a rectified linear unit (ReLU), introduce non-linearity, allowing the network to learn complex relationships.
\end{itemize}

An example of the typical architecture of a CNN is shown in Fig.~\ref{fig:traditional-convolutional-network}.

\usetikzlibrary{decorations.pathreplacing}
\usetikzlibrary{fadings}

\begin{figure*}[t!]
        \centering
        \begin{tikzpicture}
\definecolor{cnnblue}{HTML}{2c7bb6}

                \node at (0.5,-1){\begin{tabular}{c}input image\\layer \end{tabular}};
                
                \draw[fill=cnnblue,opacity=1,draw=black] (0,1) -- (1,1) -- (1,2) -- (0,2) -- (0,1);
                
                \node at (3,3){\begin{tabular}{c}convolutional layer \end{tabular}};
                
                \draw[fill=cnnblue,opacity=0.2,draw=black] (2.75,1.25) -- (3.75,1.25) -- (3.75,2.25) -- (2.75,2.25) -- (2.75,1.25);
                \draw[fill=cnnblue,opacity=0.2,draw=black] (2.5,1) -- (3.5,1) -- (3.5,2) -- (2.5,2) -- (2.5,1);
                \draw[fill=cnnblue,opacity=0.2,draw=black] (2.25,0.75) -- (3.25,0.75) -- (3.25,1.75) -- (2.25,1.75) -- (2.25,0.75);
                \draw[fill=cnnblue,opacity=0.2,draw=black] (2,0.5) -- (3,0.5) -- (3,1.5) -- (2,1.5) -- (2,0.5);
                \draw[fill=cnnblue,opacity=0.2,draw=black] (1.75,0.25) -- (2.75,0.25) -- (2.75,1.25) -- (1.75,1.25) -- (1.75,0.25);
                \draw[fill=cnnblue,opacity=0.2,draw=black] (1.5,0) -- (2.5,0) -- (2.5,1) -- (1.5,1) -- (1.5,0);
                
                \node at (4,-1){\begin{tabular}{c}pooling layer \end{tabular}};
                
                \draw[fill=cnnblue,opacity=0.2,draw=black] (5,1.25) -- (5.75,1.25) -- (5.75,2) -- (5,2) -- (5,1.25);
                \draw[fill=cnnblue,opacity=0.2,draw=black] (4.75,1) -- (5.5,1) -- (5.5,1.75) -- (4.75,1.75) -- (4.75,1);
                \draw[fill=cnnblue,opacity=0.2,draw=black] (4.5,0.75) -- (5.25,0.75) -- (5.25,1.5) -- (4.5,1.5) -- (4.5,0.75);
                \draw[fill=cnnblue,opacity=0.2,draw=black] (4.25,0.5) -- (5,0.5) -- (5,1.25) -- (4.25,1.25) -- (4.25,0.5);
                \draw[fill=cnnblue,opacity=0.2,draw=black] (4,0.25) -- (4.75,0.25) -- (4.75,1) -- (4,1) -- (4,0.25);
                \draw[fill=cnnblue,opacity=0.2,draw=black] (3.75,0) -- (4.5,0) -- (4.5,0.75) -- (3.75,0.75) -- (3.75,0);
                
                \node at (8,3){\begin{tabular}{c}convolutional layer \end{tabular}};
                
                \draw[fill=cnnblue,opacity=0.2,draw=black] (7.5,1.75) -- (8.25,1.75) -- (8.25,2.5) -- (7.5,2.5) -- (7.5,1.75);
                \draw[fill=cnnblue,opacity=0.2,draw=black] (7.25,1.5) -- (8,1.5) -- (8,2.25) -- (7.25,2.25) -- (7.25,1.5);
                \draw[fill=cnnblue,opacity=0.2,draw=black] (7,1.25) -- (7.75,1.25) -- (7.75,2) -- (7,2) -- (7,1.25);
                \draw[fill=cnnblue,opacity=0.2,draw=black] (6.75,1) -- (7.5,1) -- (7.5,1.75) -- (6.75,1.75) -- (6.75,1);
                \draw[fill=cnnblue,opacity=0.2,draw=black] (6.5,0.75) -- (7.25,0.75) -- (7.25,1.5) -- (6.5,1.5) -- (6.5,0.75);
                \draw[fill=cnnblue,opacity=0.2,draw=black] (6.25,0.5) -- (7,0.5) -- (7,1.25) -- (6.25,1.25) -- (6.25,0.5);
                \draw[fill=cnnblue,opacity=0.2,draw=black] (6,0.25) -- (6.75,0.25) -- (6.75,1) -- (6,1) -- (6,0.25);
                \draw[fill=cnnblue,opacity=0.2,draw=black] (5.75,0) -- (6.5,0) -- (6.5,0.75) -- (5.75,0.75) -- (5.75,0);
                
                \node at (8.5,-1){\begin{tabular}{c}pooling layer \end{tabular}};
                
                \draw[fill=cnnblue,opacity=0.2,draw=black] (10,1.75) -- (10.5,1.75) -- (10.5,2.25) -- (10,2.25) -- (10,1.75);
                \draw[fill=cnnblue,opacity=0.2,draw=black] (9.75,1.5) -- (10.25,1.5) -- (10.25,2) -- (9.75,2) -- (9.75,1.5);
                \draw[fill=cnnblue,opacity=0.2,draw=black] (9.5,1.25) -- (10,1.25) -- (10,1.75) -- (9.5,1.75) -- (9.5,1.25);
                \draw[fill=cnnblue,opacity=0.2,draw=black] (9.25,1) -- (9.75,1) -- (9.75,1.5) -- (9.25,1.5) -- (9.25,1);
                \draw[fill=cnnblue,opacity=0.2,draw=black] (9,0.75) -- (9.5,0.75) -- (9.5,1.25) -- (9,1.25) -- (9,0.75);
                \draw[fill=cnnblue,opacity=0.2,draw=black] (8.75,0.5) -- (9.25,0.5) -- (9.25,1) -- (8.75,1) -- (8.75,0.5);
                \draw[fill=cnnblue,opacity=0.2,draw=black] (8.5,0.25) -- (9,0.25) -- (9,0.75) -- (8.5,0.75) -- (8.5,0.25);
                \draw[fill=cnnblue,opacity=0.2,draw=black] (8.25,0) -- (8.75,0) -- (8.75,0.5) -- (8.25,0.5) -- (8.25,0);
                
                \node at (12.5,3){\begin{tabular}{c}fully connected layer\end{tabular}};
                
                \draw[fill=cnnblue,draw=black,opacity=0.5] (10.5,0) -- (11,0) -- (12.5,1.75) -- (12,1.75) -- (10.5,0);
                
                \node at (13.5,-1){\begin{tabular}{c}fully connected \\output layer \end{tabular}};
                
                \draw[fill=cnnblue,draw=black,opacity=0.5] (12.5,0.5) -- (13,0.5) -- (13.65,1.25) -- (13.15,1.25) -- (12.5,0.5);
        \end{tikzpicture}
        \caption{Architecture of a traditional CNN in which there is a sequence of convolutional and pooling layers, and after a final subsampling layer, there is a fully connected one that performs the classification or regression task. In this specific work, we removed the fully connected part, and we used only the feature extraction, which is performed up to the last pooling layer.}
        \label{fig:traditional-convolutional-network}
\end{figure*}
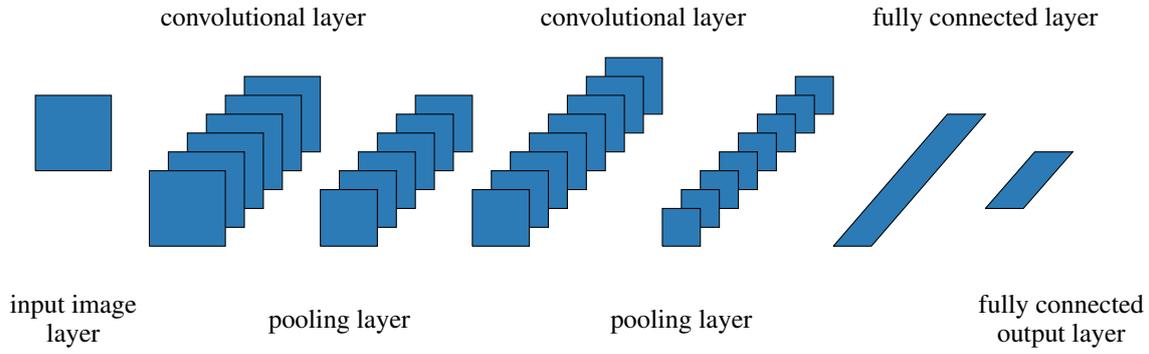

Scaling in CNNs involves adjusting various architectural dimensions to control the model complexity and computational efficiency. In particular, classical CNNs perform one of the following kinds of scaling:

\begin{itemize}
\item \textbf{Depth scaling:} Increasing the depth of a CNN involves adding more layers. Deeper networks can capture more complex features but require more computational resources.

\item \textbf{Width scaling:} Wider networks increase the number of filters in each layer. This helps in capturing more diverse features but also leads to higher computational requirements.

\item \textbf{Resolution scaling:} Resolution scaling involves adjusting the input image size. Lower resolutions reduce computational demands but may result in loss of information.
\end{itemize}

The EfficientNet network is a CNN architecture that scales up the network depth, width, and resolution in a balanced and efficient way. This method is called compound scaling, and it combines depth, width, and resolution scaling in a balanced manner, maintaining a constant ratio. It achieved state-of-the-art accuracy on ImageNet, a large-scale image dataset comprising more than 1.3 million images belonging to more than 1,000 categories. By using a pretrained EfficientNet network, we can leverage the general image processing skills learned from the ImageNet dataset and adapt them to the specific characteristics of astronomical time series. For this work, we made use only of the first part of the network that extracts the features until the last pooling layer, while we completely discarded the final fully connected layer that in EfficientNet works as classifier (see Fig.~\ref{fig:traditional-convolutional-network}).

The k-NN~\citep{Cover67} algorithm is a non-parametric supervised learning method that classifies an object based on the label of its k closest neighbors through a majority vote. In our case, we represented images by their features extracted from the EfficientNet network and measured the Euclidean distance of an image to its closest neighbor in this feature space (k=1). This distance reflects the similarity to the other images, and a large distance indicates a possible anomaly.

Since we needed a threshold to distinguish a real anomaly from something that is simply less similar, we used a stacked image (see Sec.~\ref{sec:data}) as a reference for the distance calculation. The stacked image is one that minimizes the noise and the presence of defects or spurious features with respect to the individual epochs, and it results in a sharper and cleaner reference template. We could thus safely assume that the stacked image represents the normal state of the scene and that any image that deviates significantly from it is an outlier.

In practice, as depicted in Fig.~\ref{fig:schema}, our method consists of the following steps: 

\begin{enumerate}
    \item[a.] \textbf{Input data preparation:} Time series data from astronomical observations consisting of multiple images of the same celestial object and its stacked image are collected.
    \item[b.] \textbf{Feature extraction:} A pre-trained EfficientNet network on ImageNet is used.
    EfficientNet is a CNN architecture that efficiently scales depth, width, and resolution.
    The features are collected from input images using the EfficientNet as a feature extractor.
    \item[c.] \textbf{Identification of first neighbor:} The k-NN algorithm (k=1) is applied to measure the Euclidean distance among features extracted by the EfficientNet network.
    \item[d.] \textbf{Thresholding:} The threshold for anomaly detection is identified.
    We use a stacked image as a reference for distance calculation.
    \item[e.] \textbf{Anomaly detection:} If the distance from the first neighbor is above the predefined threshold, the image is flagged as a potential outlier.
    We identify images deviating significantly from the normal state represented by the stacked image.
    \item[f.] \textbf{Output:} A list of potential outlier images identified during the anomaly detection process is produced.
\end{enumerate}
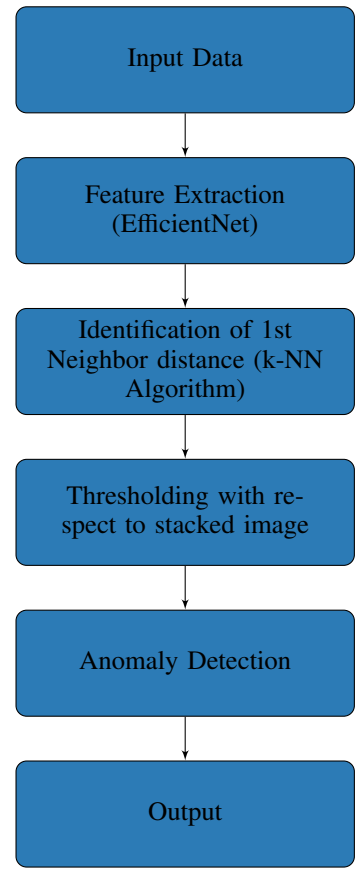
\begin{figure}[!htb]
    \centering
\tikzstyle{block} = [rectangle, draw, text width=12em, text centered, rounded corners, minimum height=4em]
\tikzstyle{line} = [draw, -latex']

\begin{tikzpicture}[node distance = 2cm, auto]
    \definecolor{bgblue}{HTML}{2c7bb6}

    \node [block, fill=bgblue,opacity=0.3,text opacity=1,draw=black] (input) {Input Data};
    \node [block, fill=bgblue,opacity=0.3,text opacity=1,draw=black, below of=input] (extract) {Feature Extraction \\ (EfficientNet)};
    \node [block, fill=bgblue,opacity=0.3,text opacity=1,draw=black, below of=extract] (knn) {Identification of 1st Neighbor distance (k-NN \\ Algorithm)};
    \node [block, fill=bgblue,opacity=0.3,text opacity=1,draw=black, below of=knn] (threshold) {Thresholding with respect to stacked image};
    \node [block, fill=bgblue,opacity=0.3,text opacity=1,draw=black, below of=threshold] (detection) {Anomaly Detection};
    \node [block, fill=bgblue,opacity=0.3,text opacity=1,draw=black, below of=detection] (output) {Output};

    \path [line] (input) -- (extract);
    \path [line] (extract) -- (knn);
    \path [line] (knn) -- (threshold);
    \path [line] (threshold) -- (detection);
    \path [line] (detection) -- (output);
\end{tikzpicture}

    \caption{Schematic description of the algorithm.}
    \label{fig:schema}
\end{figure}

\section{Experiments and results}\label{sec:results}

We decided to perform two testing campaigns, one on a larger dataset in order to have a wider case study and one on a smaller dataset, as we could then manually verify the results in a more detailed way. We decided to use as the larger dataset the Unlabel Set and as the smaller dataset the Label Set, both derived from~\cite{decicco21}.

\subsection{Large dataset}
We conducted a thorough analysis of the results obtained from applying our algorithm to the Unlabel Set dataset by~\cite{decicco21}. The dataset comprises 17,995 sources observed multiple times (Table \ref{tab:dataset}), yielding a total of 989,725 individual observations.  

As a first step, we used the same procedure of \cite{decicco19} to expunge bad epochs from the light curves, applying a sigma-clipping algorithm with a threshold of five times the value of the normalized median absolute deviation (NMAD) from the median magnitude of each source. This conservative threshold enabled us to exclude extreme outliers, preserving the most real variable events of astrophysical interest, such as AGN, supernovae, and other transients, provided the transient was not an extremely bright event in an otherwise quiescent galaxy (of which we have found no evidence in our data so far). 
This process refined the dataset to 914,156 epochs. 
Despite this initial cleaning step, the final light curves can still be affected by many photometric problems and artifacts, as discussed in Sec.~\ref{introduction}, that  may require additional ad hoc refinements to obtain fully reliable light curves. In our case, we directly applied the algorithm described in Sec.~\ref{sec:methdod}, finding 336 additional anomalous epochs.

\begin{figure}[!htb]
    \centering
\includegraphics[width=0.32\linewidth]{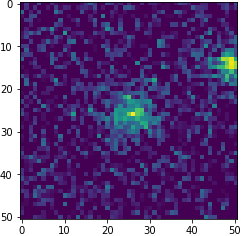}
\includegraphics[width=0.32\linewidth]{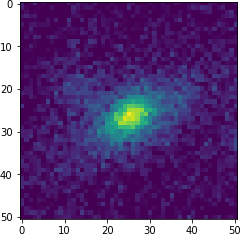}
\includegraphics[width=0.32\linewidth]{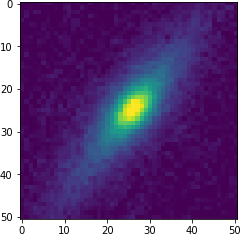}
\includegraphics[width=0.32\linewidth]{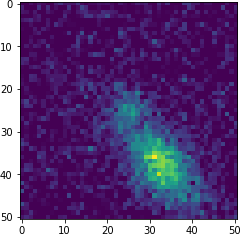}
\includegraphics[width=0.32\linewidth]{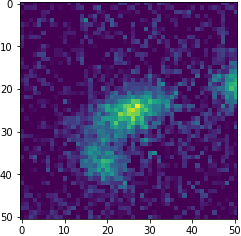}
\includegraphics[width=0.32\linewidth]{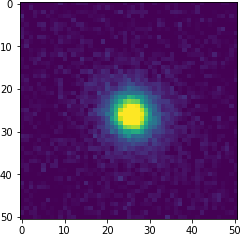}
\includegraphics[width=0.32\linewidth]{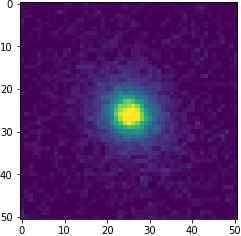}
\includegraphics[width=0.32\linewidth]{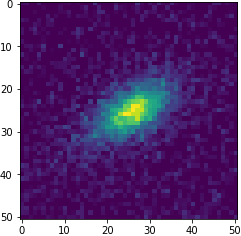}
\includegraphics[width=0.32\linewidth]{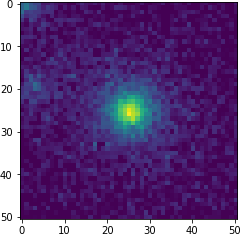}
\caption{Examples of problematic objects from epoch 32. The full set of sources flagged as problematic in epoch 32 is available in Fig.~\ref{fig:epoch32-app}.}
\label{fig:epoch32}
\end{figure}

Upon initial analysis, we observed that a specific epoch, namely epoch 32, according to the nomenclature provided in Table~\ref{tab:dataset}, exhibited 44 problematic images. A swift inspection revealed them to be images with a low S/N, as this is one of the epochs with the worst seeing values (1.18\arcsec, see Table \ref{tab:dataset}, Fig.~\ref{fig:epoch32},~\ref{fig:epoch32-app}).

\begin{figure}[!htb]
    \centering
\includegraphics[width=0.32\linewidth]{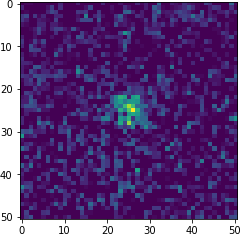}
\includegraphics[width=0.32\linewidth]{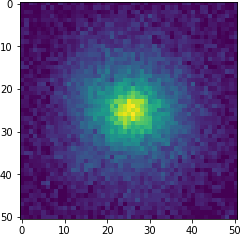} 

\includegraphics[width=0.32\linewidth]{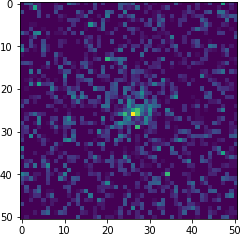}
\includegraphics[width=0.32\linewidth]{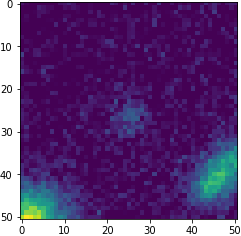}
\caption{Sources flagged as outliers in epoch 34 due to low S/N ratio with regard to the reference epoch.}
\label{fig:epoch34}
\end{figure}

On four occasions, epoch 34 produced outliers linked to a low S/N ratio (see Fig.~\ref{fig:epoch34}). Epoch 34 actually has a worse seeing (i.e., 1.28\arcsec) than Epoch 32.

\begin{figure}[!htb]
\centering

\includegraphics[width=0.32\linewidth]{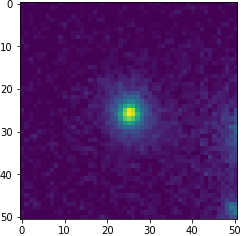}
\includegraphics[width=0.32\linewidth]{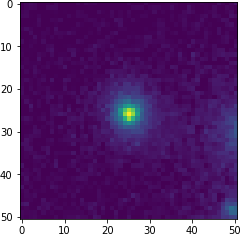}
\includegraphics[width=0.32\linewidth]{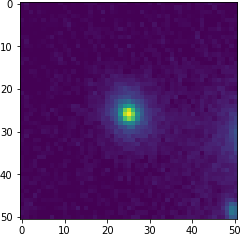}
\includegraphics[width=0.32\linewidth]{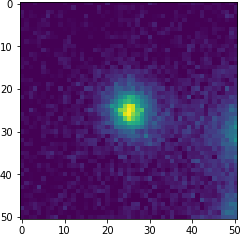}
\includegraphics[width=0.32\linewidth]{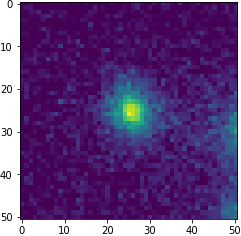}
\includegraphics[width=0.32\linewidth]{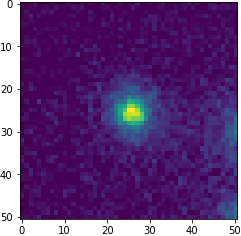}
\includegraphics[width=0.32\linewidth]{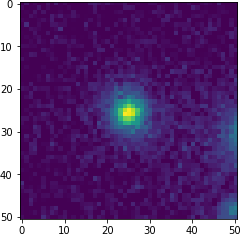}
\includegraphics[width=0.32\linewidth]{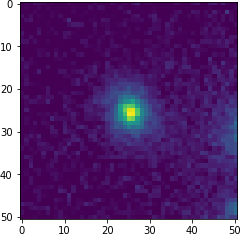}
\includegraphics[width=0.32\linewidth]{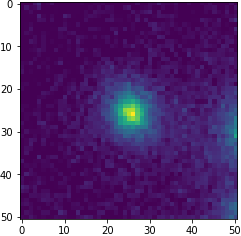}
\includegraphics[width=0.4455\linewidth]{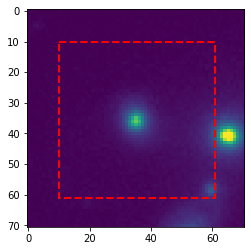}
\caption{Thumbnails of the flagged epochs for source 23560 (including epoch 32, which was excluded from Fig.~\ref{fig:epoch32-app}). The last panel is a broader view of the stacked image showing the nearby contaminant; the size of the previous thumbnails is shown by the red dashed square.}
\label{fig:obj_23560}
\end{figure}

Another noteworthy case concerns source 23560, where nine out of 54 epochs were identified as anomalous, likely due to the contamination from a bright neighbor (see Fig.~\ref{fig:obj_23560}). In fact, source 23560 has a bright companion that can be seen in the bottom panel of Fig.~\ref{fig:obj_23560}, so we are in the presence of a configuration that, due also to the fluctuation in the point spread function, affects the companion outskirts that contaminate this source and is spotted by our algorithm.

\begin{figure}[!htb]
    \centering
\includegraphics[width=0.32\linewidth]{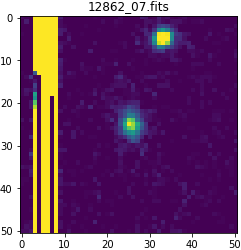}
\includegraphics[width=0.32\linewidth]{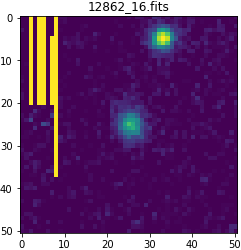}
\includegraphics[width=0.32\linewidth]{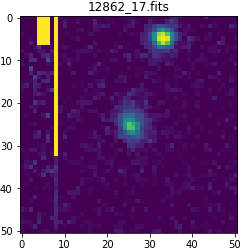}

\includegraphics[width=0.32\linewidth]{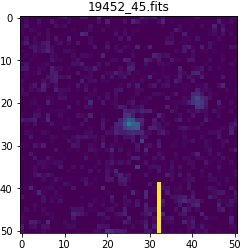}

\includegraphics[width=0.32\linewidth]{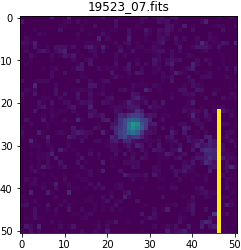}
\includegraphics[width=0.32\linewidth]{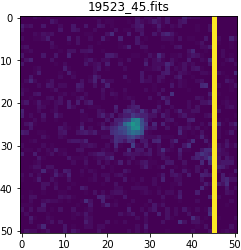}

\includegraphics[width=0.32\linewidth]{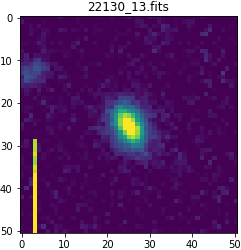}

\includegraphics[width=0.32\linewidth]{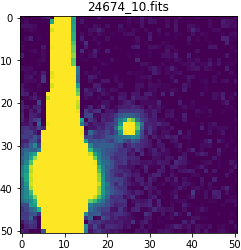}
\includegraphics[width=0.32\linewidth]{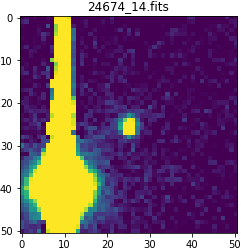}
\includegraphics[width=0.32\linewidth]{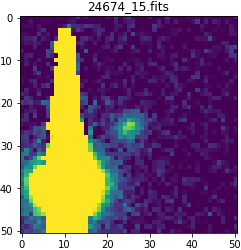}
\includegraphics[width=0.32\linewidth]{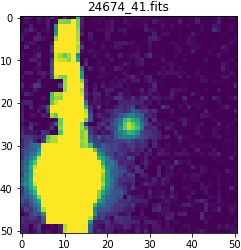}
\includegraphics[width=0.32\linewidth]{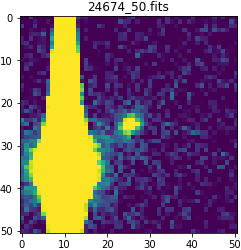}
\caption{Objects with a bright source nearby. Each row corresponds to a different object with some flagged epochs, with the exception of the last two lines, as they correspond to the same object.}
\label{fig:brightSource}
\end{figure}

Twelve additional epochs showed contamination because of defects from electronics or saturated neighbors. Even when the central object flux does not seem directly affected, the algorithm background estimation might be distorted (Fig.~\ref{fig:brightSource}).

\begin{figure}[!htb]
    \centering
\includegraphics[width=0.32\linewidth]{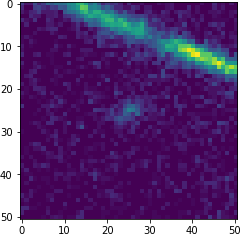}
\includegraphics[width=0.32\linewidth]{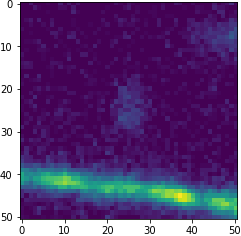}
\includegraphics[width=0.32\linewidth]{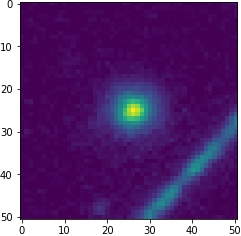}
\includegraphics[width=0.32\linewidth]{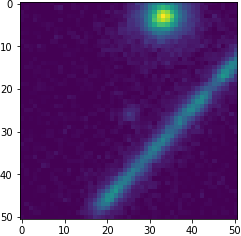}
\includegraphics[width=0.32\linewidth]{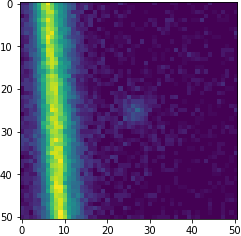}
\includegraphics[width=0.32\linewidth]{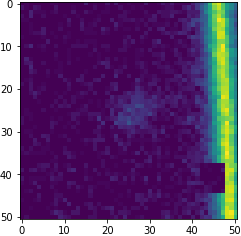}
\includegraphics[width=0.32\linewidth]{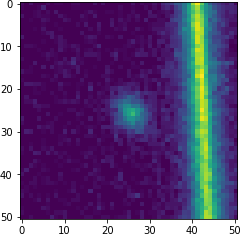}
\includegraphics[width=0.32\linewidth]{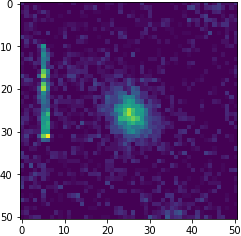}
\includegraphics[width=0.32\linewidth]{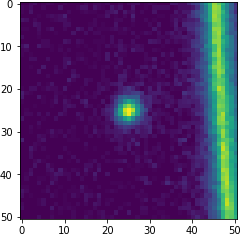}
\caption{Some examples of objects with tracks. (See 
Fig.~\ref{fig:track1} for the whole sample.)}
    \label{fig:track}
\end{figure}

Additionally, we identified 185 instances of astronomical tracks (asteroids, satellites, etc.) that, while not directly influencing the central source flux, could distort background estimation and thus the magnitude determination (Figs.~\ref{fig:track},~\ref{fig:track1}).

\begin{figure}[!htb]
    \centering
\includegraphics[width=0.32\linewidth]{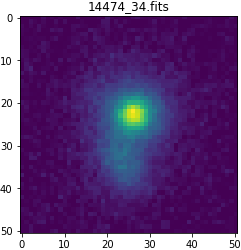}
\includegraphics[width=0.32\linewidth]{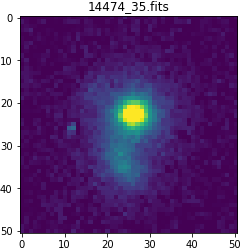}
\includegraphics[width=0.32\linewidth]{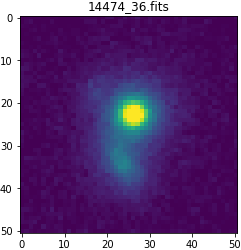}
\includegraphics[width=0.32\linewidth]{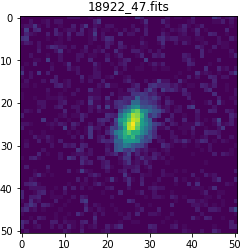}
\includegraphics[width=0.32\linewidth]{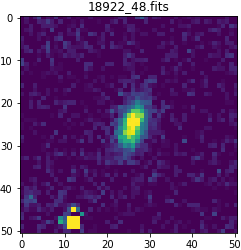}
\includegraphics[width=0.32\linewidth]{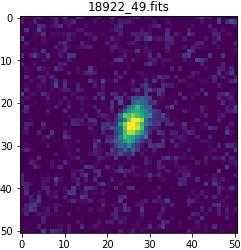}
\includegraphics[width=0.32\linewidth]{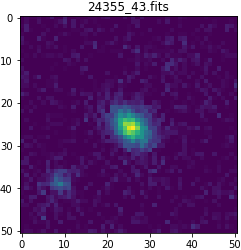}
\includegraphics[width=0.32\linewidth]{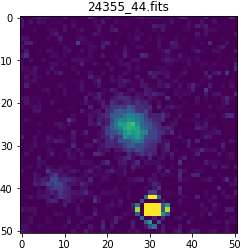}
\includegraphics[width=0.32\linewidth]{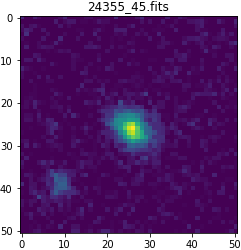}
\caption{Sources with an additional object that appears nearby. For each object, the epoch flagged by the method is shown in the central column. The epoch before and the one after are also shown in the first and last column, respectively. For the first object, the new object detection corresponds to the faint object located approximately at (10,25). }
\label{fig:newObj1}
\end{figure}

In three cases, an epoch was flagged due to the appearance of a secondary object in a single epoch (Fig.~\ref{fig:newObj1}). While this is an intriguing collateral study, it is unlikely to impact variability measurements; nonetheless, the algorithm correctly identified them as outliers.

\begin{figure}[!htbp]
     \centering
\includegraphics[width=.475\linewidth]{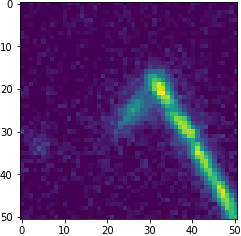}
\includegraphics[width=.475\linewidth]{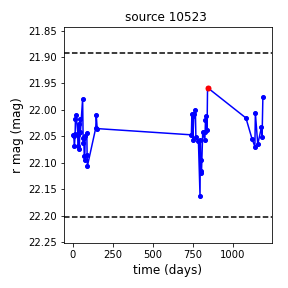}
\includegraphics[width=.475\linewidth]{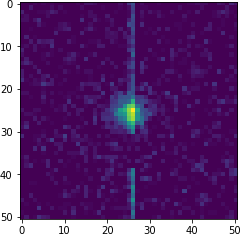}     
\includegraphics[width=.475\linewidth]{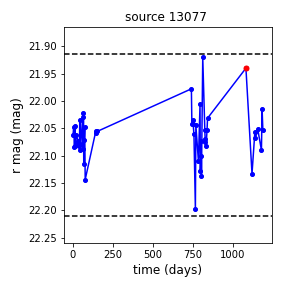}  
\includegraphics[width=.475\linewidth]{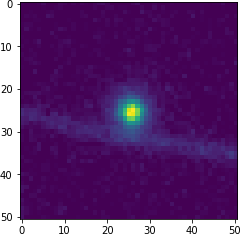}
\includegraphics[width=.475\linewidth]{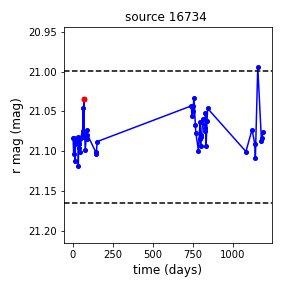}
\includegraphics[width=.475\linewidth]{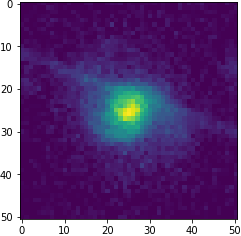}   
\includegraphics[width=.475\linewidth]{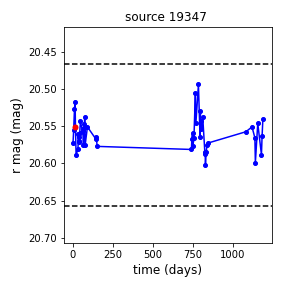}
\includegraphics[width=.475\linewidth]{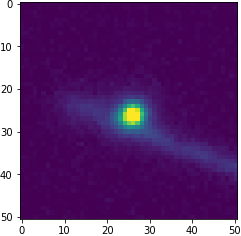}
\includegraphics[width=.475\linewidth]{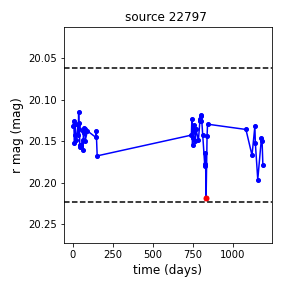}
\caption{Some of the problematic objects identified by our method. In most cases, if a track appears, it is closer to the object with respect to the aperture photometry, so it slightly affects the flux measurement. In the left column, one can see the image as identified by our algorithm, while in the right we report the light curve with the flagged epoch marked with red. The horizontal lines represent the threshold of the sigma clip. (See Fig.~\ref{fig:problematic-app} and~\ref{fig:problematic-lc-app} for the whole sample.)}
    \label{fig:problematic}
\end{figure}

The most significant finding pertains to 52 events where source measurements were undoubtedly influenced to some extent by tracks or other issues (Figs.~\ref{fig:problematic}, \ref{fig:problematic-app}, \ref{fig:problematic-lc-app}). These epochs should be excluded from any further study involving these objects. In particular, we would like to emphasize that, with some exceptions, those epochs could not be spotted with a different threshold of the sigma clip since they lie in the core of the distribution of the light curve despite being spoiled. 
There are some cases, such as the last line of Fig.~\ref{fig:problematic}, where although there is an additional flux coming from a track, the measured flux from the pipeline is lower. In this specific case, we observed that due to the track, the pipeline provided a wrong estimate of the centroid of the object, and therefore the aperture magnitude  does not contain all the flux of the actual source. Since our method does not rely on centroid measurements but makes use of the original images, it is able to identify these outliers anyway.

\begin{figure}[!htb]
    \centering
\includegraphics[width=0.32\linewidth]{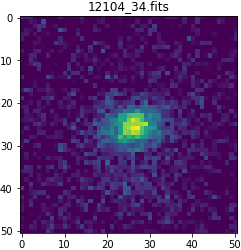}
\includegraphics[width=0.32\linewidth]{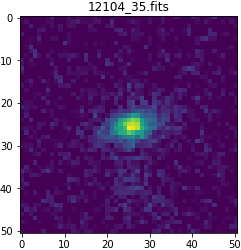}
\includegraphics[width=0.32\linewidth]{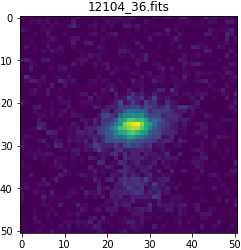}
\hfill
\includegraphics[width=0.32\linewidth]{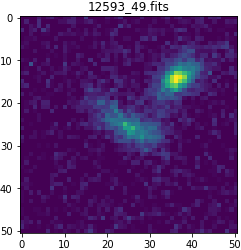}
\includegraphics[width=0.32\linewidth]{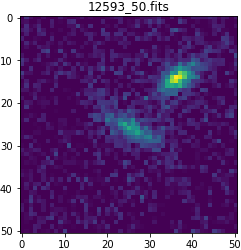}
\includegraphics[width=0.32\linewidth]{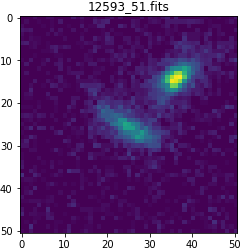}

\includegraphics[width=0.32\linewidth]{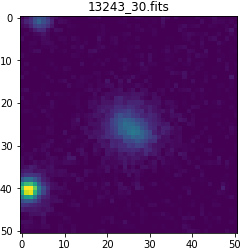}
\includegraphics[width=0.32\linewidth]{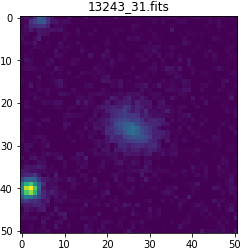}
\includegraphics[width=0.32\linewidth]{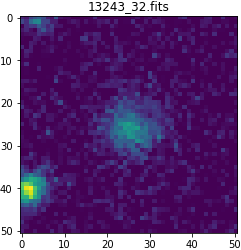}
\caption{Three of the objects flagged as anomalies for which we do not have a proper motivation. Some of them can still be low S/N objects. For each source, the epoch flagged is the central one, while on the left and on the right, we have the previous and the next epoch, respectively. (See Figs.~\ref{fig:boh1-app} and~\ref{fig:boh2} for the whole sample.)}
\label{fig:boh1}
\end{figure}

Finally there remain 27 epochs for which, despite being flagged as outliers, a clear interpretation could not be found. They represent approximately 8\% of the flagged sources (that corresponds to 0.003\% of the whole set of images presented to the algorithm) and could be considered as false positives identified by our algorithm (Fig.~\ref{fig:boh1}, \ref{fig:boh1-app}, and~\ref{fig:boh2}).
However, we observed that a significant portion (approximately 22\% of these cases) belonged to one of the two epochs with the best seeing conditions (epochs 14 and 6, with seeing of 0.5\arcsec and 0.58\arcsec\ respectively; see Table.\ref{tab:dataset}). This suggests that given the exceptionally good seeing conditions, these objects exhibit even greater dissimilarities than others when compared to the stacked image.
However, considering the minimal information loss (0.003\% relative to the original dataset), we find this result to be more than acceptable.

Finally, we performed a test on all 989,725 individual observations, including epochs rejected by the sigma-clipping criterion, which resulted in 1,715 flagged images. A large part of these are already included in the $\sim85,000$ objects discarded by the sigma clipping, while the majority of the remaining objects are the same ones selected by our approach after applying the sigma clipping. It is clear, however, that our algorithm is not capable of reproducing the sigma-clipping results, and therefore it is better to use it after the preliminary cleaning procedure as a further refinement in order to flag and eliminate most of the remaining outliers.

\subsection{Small dataset}

We then transitioned to utilizing the Label Set dataset. This smaller dataset enabled us to visually explore data filtered by sigma clipping. 
It was derived by \cite{decicco21} and consists of all the objects in the field for which a classification of the sources, in terms of star, galaxy, and AGN, is available. The size of this dataset made it an ideal case for testing for any possible issue with our new algorithm. Our method did not identify all the anomalies in this dataset, comprising 2,675 sources and a total of 147,125 individual observations. Post sigma clipping, this number decreased to 140,920, with 6,205 epochs eliminated by the sigma clip.

By executing our algorithm on the complete Label Set dataset, we identified 289 anomalies, of which 202 were already flagged by the sigma clip. For the anomalies not identified by the sigma clip, a significant number was found to be tracks, in some cases overlapping the source. 

Our focus then shifted to the 6,003 sources that were clipped but not identified by our algorithm. Notably, these sources did not exhibit images entirely distorted by artifacts that we know are present based on the unfiltered dataset, suggesting both methods effectively detect and filter such image-related issues.

However, there are some cases where something was overlooked, although we expected our algorithm to easily spot the epoch as anomalous. We explore two examples of the few cases where our algorithm did not identify the problem.

\begin{figure}[!hbt]
    \centering
\includegraphics[width=0.99\linewidth]{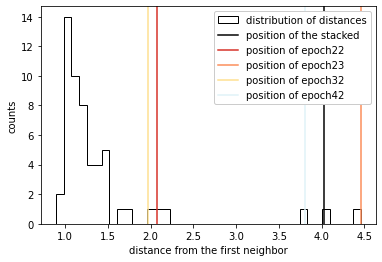}
\includegraphics[width=0.4\linewidth]{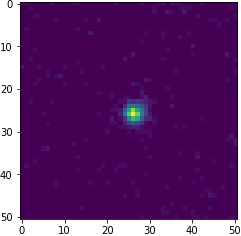}

\includegraphics[width=0.24\linewidth]{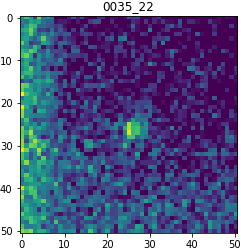}
\includegraphics[width=0.24\linewidth]{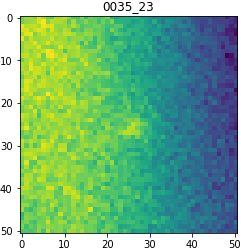}
\includegraphics[width=0.24\linewidth]{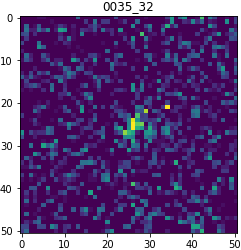}
\includegraphics[width=0.24\linewidth]{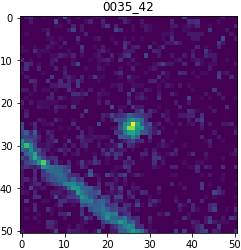}

\caption{Source 0035. First row: Histogram of the distances from the stacked image. Second row: Stacked image. Last row: Epochs 22, 23, 32, and 42.}
\label{fig:0035}
\end{figure}

For source 0035, by manually inspecting the light curve, we identified the following issue:
In epochs 22 and 23, there is a large contamination coming from a defect of the image on the left, while in epoch 42 there is a track. In addition, there are epochs with a lower S/N (for instance epoch 32). The algorithm spotted epoch 23, where the contamination is bigger, but completely missed the two other occurrences. By looking at the histogram of the distances (see Fig.~\ref{fig:0035}) as computed by our algorithm, we noticed that all the problematic epochs were in fact detected as outliers of the distribution but are simply not different enough to meet our exclusion criteria.

\begin{figure}[!ht]
    \centering
\includegraphics[width=0.99\linewidth]{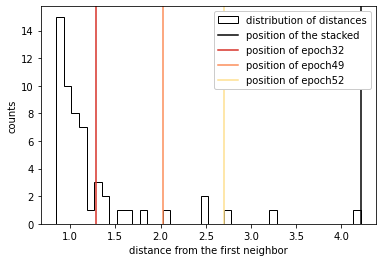}
\includegraphics[width=0.4\linewidth]{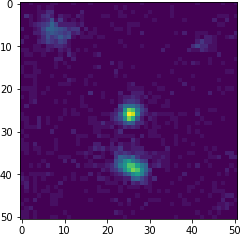}

\includegraphics[width=0.24\linewidth]{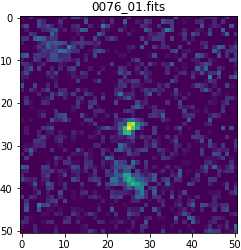}
\includegraphics[width=0.24\linewidth]{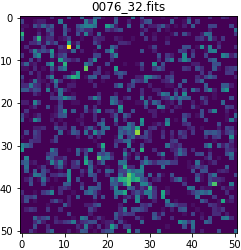}
\includegraphics[width=0.24\linewidth]{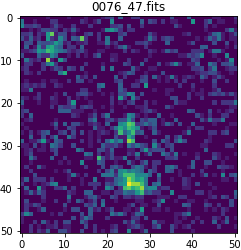}
\includegraphics[width=0.24\linewidth]{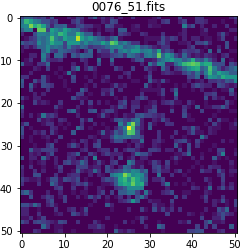}
\caption{Source 0076. First row: Histogram of the distances. Second row: Stacked image. Last row: Epochs 1, 32, 47, and 51.}
\label{fig:0076}
\end{figure}

Source 0076 exhibits a track in epoch 51. In Fig.~\ref{fig:0076}, we show two more epochs with epochs 51 and 32 for comparison. From the histogram, it appears that the track, although being anomalous, is considered more similar to the stacked image; we spotted a similar behavior in other sources. In all of these cases, the source with the track is outside the distribution of standard epochs, but the stacked image is still above it. This effect could be mitigated by considering a "weirdness" threshold. Regardless of the position of the stacked image, the image was flagged as to be rejected, and clearly this, on one hand, reduces the number of wrong epochs not flagged, but, on the other hand, it increases the number of good epochs rejected. Although we added this possibility to the code, we consider that performing the sigma clip before implementing our algorithm provides the best results.

\section{Discussion and conclusion}\label{sec:conclusion}




In this study, we propose a novel method for detecting outliers in astronomical time series, using VST monitoring campaigns of the COSMOS field as a test bed.
Astronomical time series, crucial for understanding celestial phenomena, present challenges such as noise, gaps, contaminants, and artifacts. Our approach, which falls under so-called transfer learning methods, combines a deep neural network, specifically an EfficientNet network trained on ImageNet, with a k-NN algorithm. The EfficientNet network serves as a feature extractor, and the k-NN algorithm measures the distance in the feature space to identify potential outliers. Anomalies are flagged based on the distance from the first neighbor and comparison with the same distance obtained using the stacked image, representing the normal state of the scene that has, by construction, a higher S/N and hence represents an outlier by definition.

Our experiments were separated into two cases extracted from the same dataset, one involving a larger dataset (Unlabel Set) and one involving a smaller dataset (Label Set), in order to verify the effectiveness of our method in detecting various anomalies, including a low S/N ratio, contamination from neighboring sources, data reduction errors, bright objects affecting background estimation, and tracks. The performance of the algorithm was evaluated in comparison to a classic sigma-clipping procedure commonly used in astronomical data analysis.

In the case of the larger dataset (Unlabel Set), our method identified anomalies that were missed by the sigma-clipping algorithm, including instances of a low S/N ratio, contamination, and tracks. Additionally, it successfully flagged epochs affected by bright saturated objects, data reduction errors, and off-center tracks (regarding the latter aspects, the method seems to be very efficient). Although there were some epochs that were flagged by the method for which we do not have a proper explanation, even when considering all of them as false positives, they are negligible in number with respect to the whole dataset. 

In the case of the smaller dataset (Label Set), we did the opposite by looking at the anomalies identified by the sigma clip that were not flagged by our algorithm. 
In most cases, the distribution of the distances showed that the epochs that should clearly be removed visually, although deviating from the distribution, are below the threshold identified by the stacked image. This suggests that both of the analyses need to be performed, that is, the sigma clipping first and then a refinement with the method proposed in this work.

One key aspect of our method is that unlike traditional approaches, such as sigma clipping, that focus on identifying outliers in a limited parameter space (e.g., extreme magnitude variations in the light curve), our method examines the overall structure of the images across the time series, considering the relationships and patterns in the feature space. This is particularly evident in the light curves presented in the appendix (see Fig.~\ref{fig:problematic-lc-app}), where anomalies flagged by our method go beyond the typical peak-detection scenarios. Peaks alone do not capture the diverse range of anomalies present in astronomical time series, and our method provides a more comprehensive approach to outlier detection. In substance, a simple adjustment of the sigma-clipping parameters would not suffice to capture the nuanced anomalies identified by our method. This highlights the complementary nature of our approach, offering a distinct and valuable perspective in the realm of artifact detection in astronomical time series.

The presence of defected epochs in astronomical time series can significantly impact statistical indicators commonly derived from light curves. One notable example is the pair slope, a metric used for instance in \cite{decicco21}. The pair slope measures the trend of variability between consecutive data points in a light curve, providing insights into the underlying astrophysical processes. However, the accuracy of such indicators is compromised when defected epochs are present, and these epochs may hamper the ability to detect and characterize a variety of important astrophysical events, such as tidal disruption events, supernovae, AGN, and blazars. Our method, by effectively identifying and flagging these anomalies, contributes to the preservation of the integrity of statistical measures derived from light curves, ensuring that the results are more reliable and reflective of the true astrophysical phenomena.

As a further development in the future, we plan to explore different pretrained networks, such as vision transformers \citealt{Dosovitskiy2020}. Additionally, a possible improvement could be found by exploring adaptive thresholding methods based on the distribution of the distance meant to work with the threshold determined by the stacked image in order to make the algorithm more robust to variations in data characteristics.

In addition to the effectiveness of the proposed method in detecting outliers in astronomical time series, it is noteworthy to discuss the computational efficiency and parallelization potential. The nature of the algorithm lends itself to an ``embarrassingly'' parallel paradigm. Each time series is entirely independent from the others, allowing for parallel execution on separate computational nodes. This parallelization capability makes the method well suited for deployment on high-performance computing clusters or distributed computing environments.

During our testing phase, the algorithm demonstrated impressive efficiency even on a machine without a GPU -- we utilized an Intel\textsuperscript{\textregistered}
 Core\textsuperscript{\texttrademark} i9-10980XE CPU @ 3.00GHz.\footnote{\url{https://ark.intel.com/content/www/us/en/ark/products/198017/intel-core-i9-10980xe-extreme-edition-processor-24-75m-cache-3-00-ghz.html}} On average, processing each time series consisting of a maximum of 54 epochs took less than 1 second. It is important to highlight that this timing was achieved on a CPU, and the use of a GPU is expected to further accelerate the processing time.

The scalability and adaptability of the algorithm to parallel processing environments make it a promising solution to efficiently handle large-scale astronomical datasets, such as the ones that will be obtained from the LSST. Future implementations may leverage GPU resources to achieve even faster processing times, which is particularly beneficial when dealing with extensive datasets generated by modern astronomical surveys.

\section*{Acknowledgements}

DD acknowledges PON R\&I 2021, CUP E65F21002880003. DD and MP also acknowledge the financial contribution from PRIN-MIUR 2022 and from the Timedomes grant within the ``INAF 2023 Finanziamento della Ricerca Fondamentale''. MB and SC acknoweldge the ASI-INAF TI agreement, 2018-23-HH.0 "Attività scientifica per la missione Euclid - fase D"

Topcat \citep{Taylor05} and STILTS \citep{Taylor06} have been used for this work.
Some of the resources from \cite{Stutz2022} has been used for this work.
The k-NN used in this work is part of the Scikit package \citep{Pedregosa11}.



\bibliographystyle{aa} 
\bibliography{main}

\onecolumn
\appendix
\section{Complete figures}
\begin{figure*}[!hb]
    \centering
\includegraphics[width=0.138\linewidth]{epoch32/10216_32.png}
\includegraphics[width=0.138\linewidth]{epoch32/10395_32.png}
\includegraphics[width=0.138\linewidth]{epoch32/10600_32.png}
\includegraphics[width=0.138\linewidth]{epoch32/12294_32.png}
\includegraphics[width=0.138\linewidth]{epoch32/12729_32.png}
\includegraphics[width=0.138\linewidth]{epoch32/13117_32.png}
\includegraphics[width=0.138\linewidth]{epoch32/13212_32.png}
\includegraphics[width=0.138\linewidth]{epoch32/13493_32.png}
\includegraphics[width=0.138\linewidth]{epoch32/13523_32.png}
\includegraphics[width=0.138\linewidth]{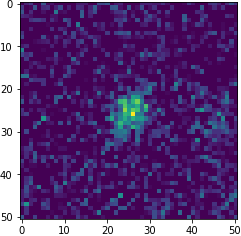}
\includegraphics[width=0.138\linewidth]{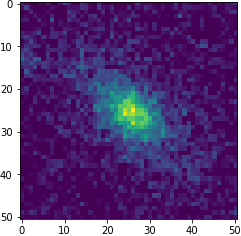}
\includegraphics[width=0.138\linewidth]{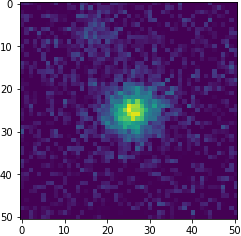}
\includegraphics[width=0.138\linewidth]{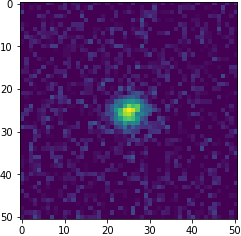}
\includegraphics[width=0.138\linewidth]{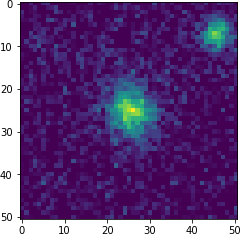}
\includegraphics[width=0.138\linewidth]{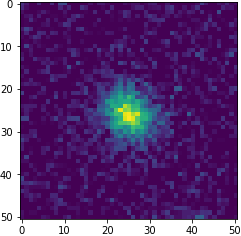}
\includegraphics[width=0.138\linewidth]{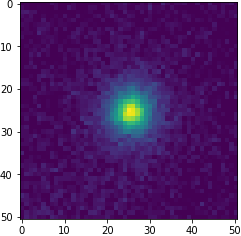}
\includegraphics[width=0.138\linewidth]{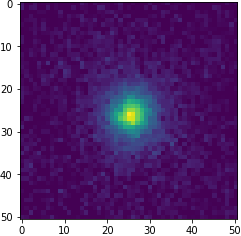}
\includegraphics[width=0.138\linewidth]{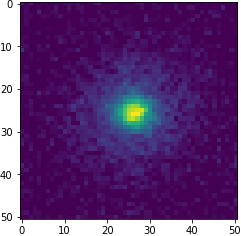}
\includegraphics[width=0.138\linewidth]{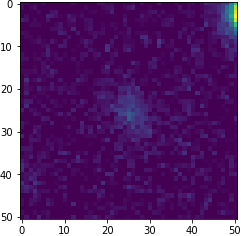}
\includegraphics[width=0.138\linewidth]{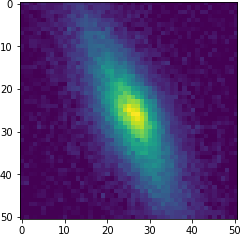}
\includegraphics[width=0.138\linewidth]{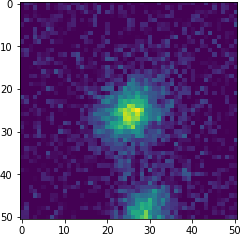}
\includegraphics[width=0.138\linewidth]{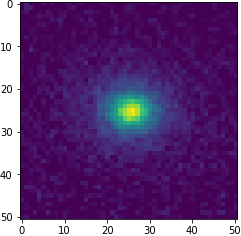}
\includegraphics[width=0.138\linewidth]{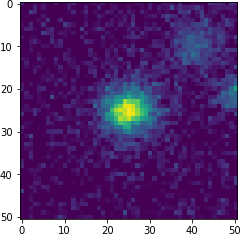}
\includegraphics[width=0.138\linewidth]{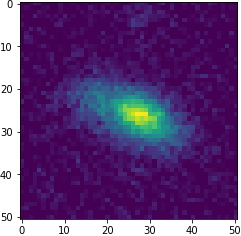}
\includegraphics[width=0.138\linewidth]{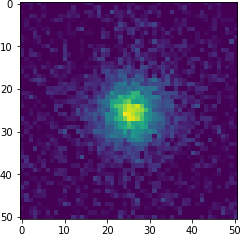}
\includegraphics[width=0.138\linewidth]{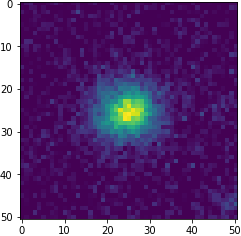} 
\includegraphics[width=0.138\linewidth]{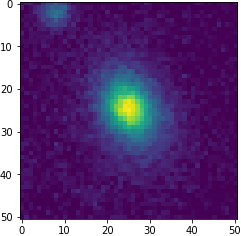}
\includegraphics[width=0.138\linewidth]{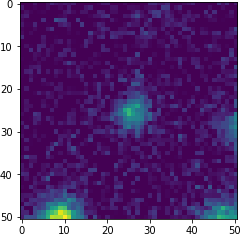}
\includegraphics[width=0.138\linewidth]{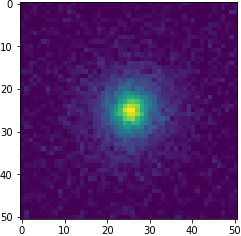}
\includegraphics[width=0.138\linewidth]{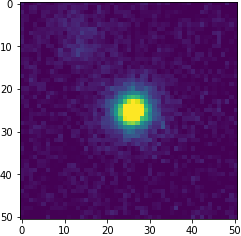}
\includegraphics[width=0.138\linewidth]{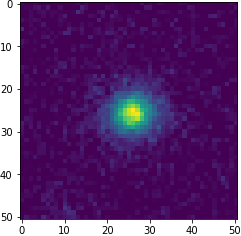}
\includegraphics[width=0.138\linewidth]{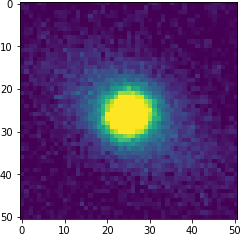}
\includegraphics[width=0.138\linewidth]{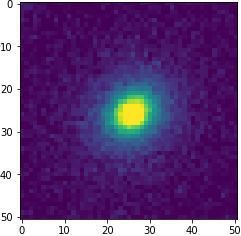}
\includegraphics[width=0.138\linewidth]{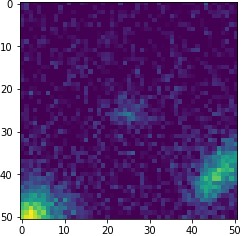} 
\includegraphics[width=0.138\linewidth]{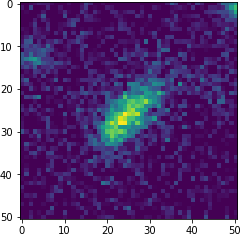}
\includegraphics[width=0.138\linewidth]{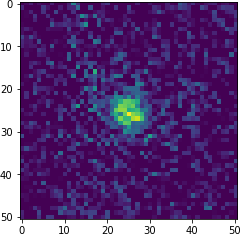}
\includegraphics[width=0.138\linewidth]{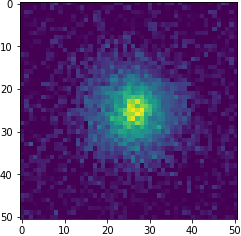}
\includegraphics[width=0.138\linewidth]{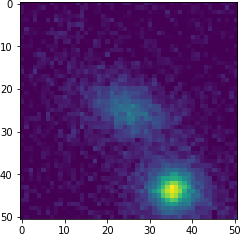}
\includegraphics[width=0.138\linewidth]{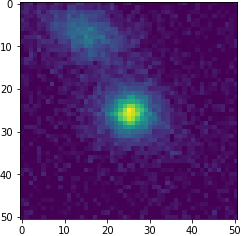}
\includegraphics[width=0.138\linewidth]{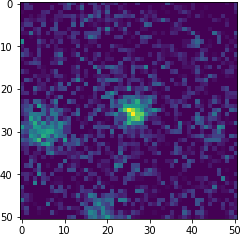}
\includegraphics[width=0.138\linewidth]{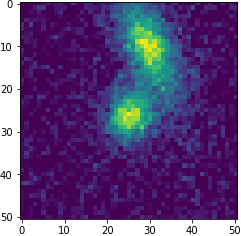}
\includegraphics[width=0.138\linewidth]{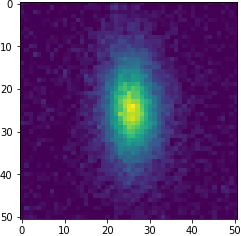}
\includegraphics[width=0.138\linewidth]{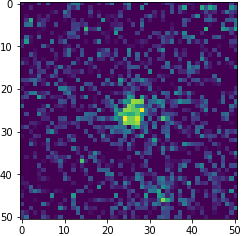}
\includegraphics[width=0.138\linewidth]{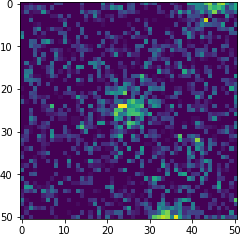} 
\caption{Full set of sources flagged as problematic in epoch 32.}
\label{fig:epoch32-app}
\end{figure*}

\begin{figure*}[!hp]

    \centering
\includegraphics[width=0.137\linewidth]{track/10686_37.png}
\includegraphics[width=0.137\linewidth]{track/11045_09.png}
\includegraphics[width=0.137\linewidth]{track/11117_46.png}
\includegraphics[width=0.137\linewidth]{track/11134_46.png}
\includegraphics[width=0.137\linewidth]{track/11506_02.png}
\includegraphics[width=0.137\linewidth]{track/11715_02.png}
\includegraphics[width=0.137\linewidth]{track/11900_02.png}
\includegraphics[width=0.137\linewidth]{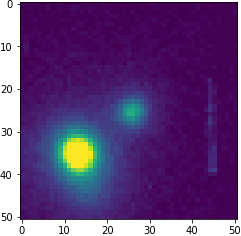}
\includegraphics[width=0.137\linewidth]{track/11940_22.png}
\includegraphics[width=0.137\linewidth]{track/13010_02.png}
\includegraphics[width=0.137\linewidth]{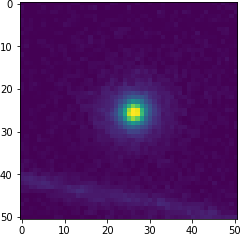}
\includegraphics[width=0.137\linewidth]{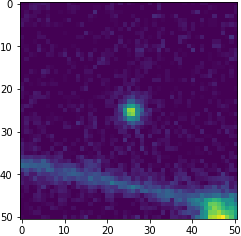}
\includegraphics[width=0.137\linewidth]{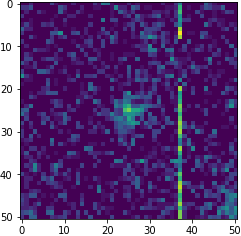}
\includegraphics[width=0.137\linewidth]{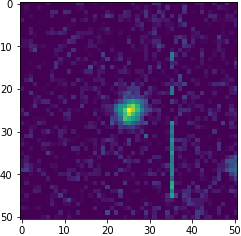}
\includegraphics[width=0.137\linewidth]{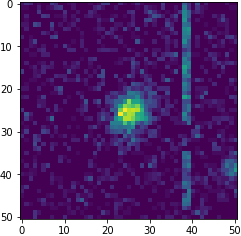}
\includegraphics[width=0.137\linewidth]{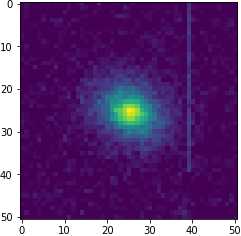}
\includegraphics[width=0.137\linewidth]{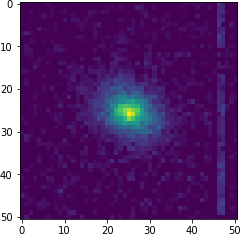}
\includegraphics[width=0.137\linewidth]{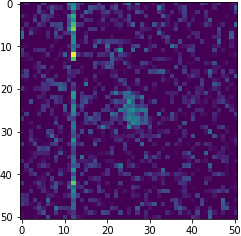}
\includegraphics[width=0.137\linewidth]{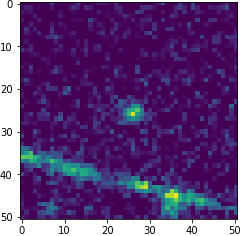}
\includegraphics[width=0.137\linewidth]{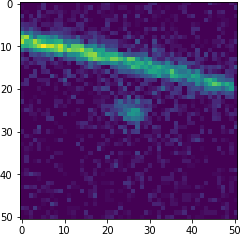}
\includegraphics[width=0.137\linewidth]{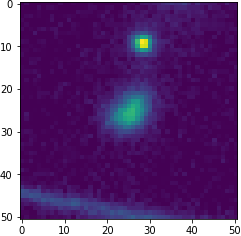}
\includegraphics[width=0.137\linewidth]{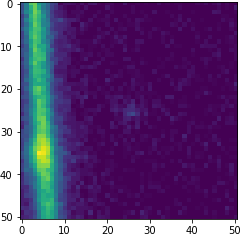}
\includegraphics[width=0.137\linewidth]{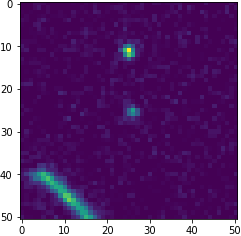}
\includegraphics[width=0.137\linewidth]{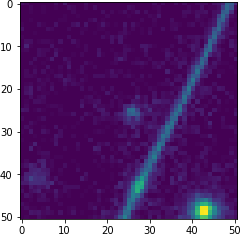}
\includegraphics[width=0.137\linewidth]{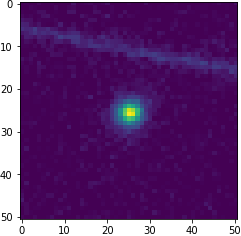}
\includegraphics[width=0.137\linewidth]{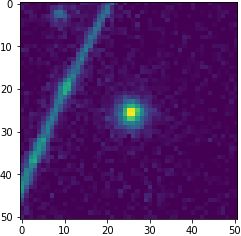}
\includegraphics[width=0.137\linewidth]{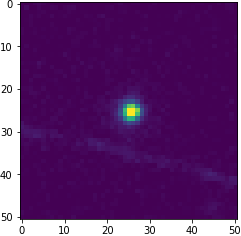}
\includegraphics[width=0.137\linewidth]{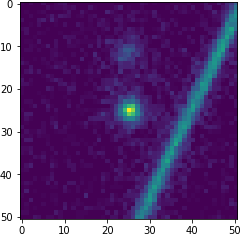}
\includegraphics[width=0.137\linewidth]{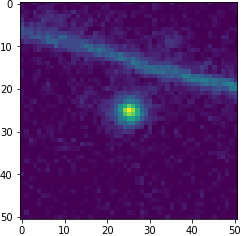}
\includegraphics[width=0.137\linewidth]{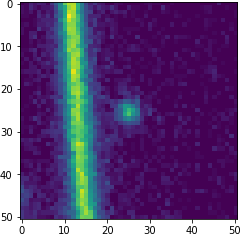}
\includegraphics[width=0.137\linewidth]{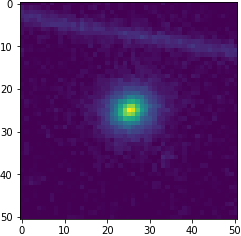}
\includegraphics[width=0.137\linewidth]{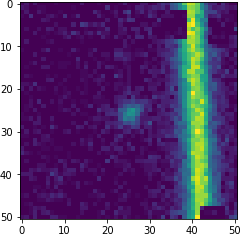}
\includegraphics[width=0.137\linewidth]{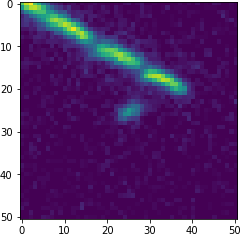}
\includegraphics[width=0.137\linewidth]{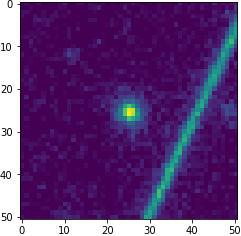}
\includegraphics[width=0.137\linewidth]{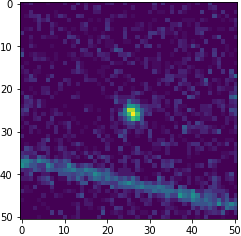}
\includegraphics[width=0.137\linewidth]{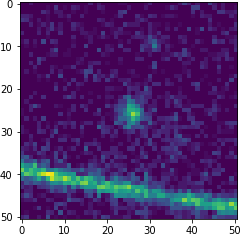}
\includegraphics[width=0.137\linewidth]{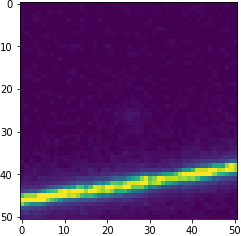}
\includegraphics[width=0.137\linewidth]{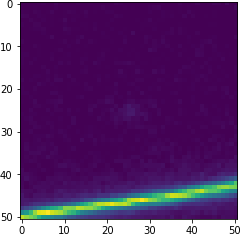}
\includegraphics[width=0.137\linewidth]{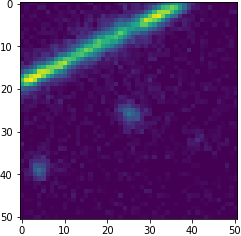}
\includegraphics[width=0.137\linewidth]{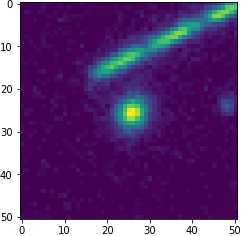}
\includegraphics[width=0.137\linewidth]{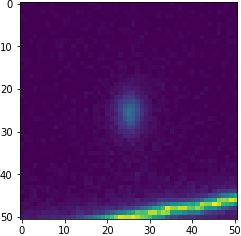}
\includegraphics[width=0.137\linewidth]{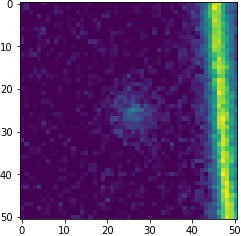}
\includegraphics[width=0.137\linewidth]{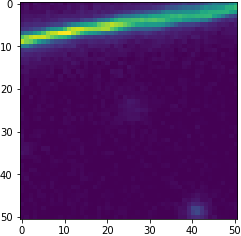}
\includegraphics[width=0.137\linewidth]{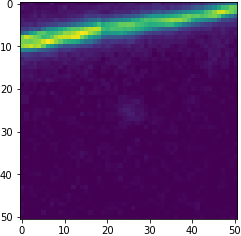}
\includegraphics[width=0.137\linewidth]{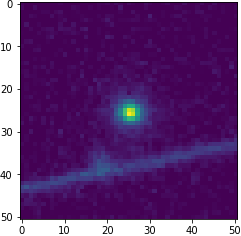}
\includegraphics[width=0.137\linewidth]{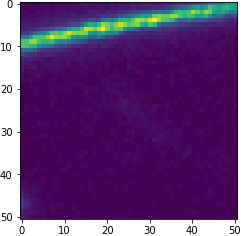}
\includegraphics[width=0.137\linewidth]{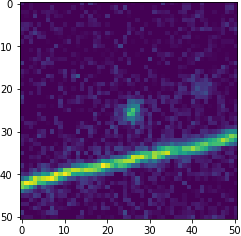}
\includegraphics[width=0.137\linewidth]{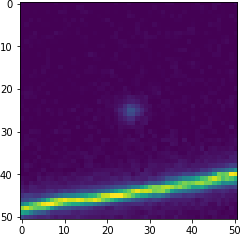}
\includegraphics[width=0.137\linewidth]{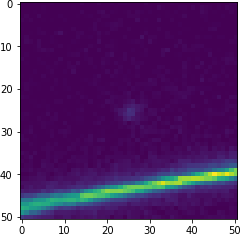}
\includegraphics[width=0.137\linewidth]{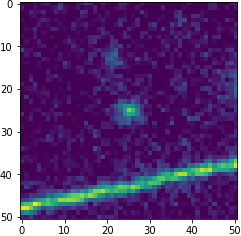}
\includegraphics[width=0.137\linewidth]{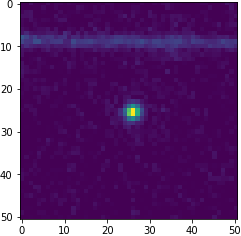}
\includegraphics[width=0.137\linewidth]{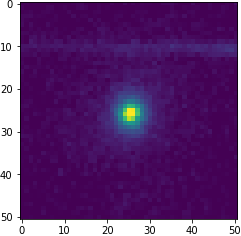}
\includegraphics[width=0.137\linewidth]{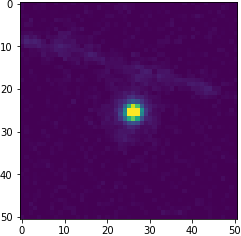}
\includegraphics[width=0.137\linewidth]{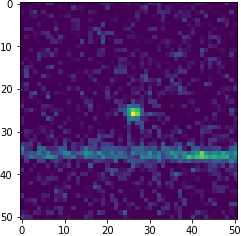}
\includegraphics[width=0.137\linewidth]{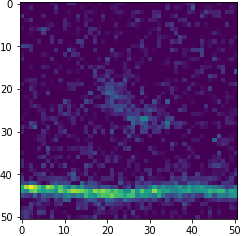}
\includegraphics[width=0.137\linewidth]{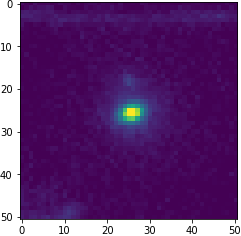}
\caption{Objects with track.}
\ContinuedFloat
    \label{fig:track1}
\end{figure*}

\begin{figure*}[!hp]
\ContinuedFloat
    \centering
\includegraphics[width=0.137\linewidth]{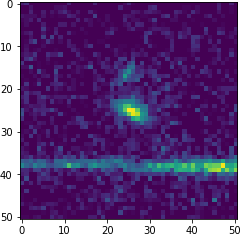}
\includegraphics[width=0.137\linewidth]{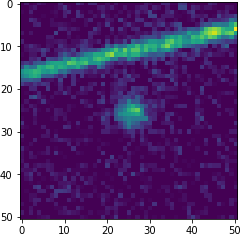}
\includegraphics[width=0.137\linewidth]{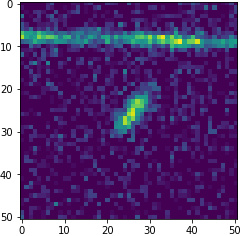}
\includegraphics[width=0.137\linewidth]{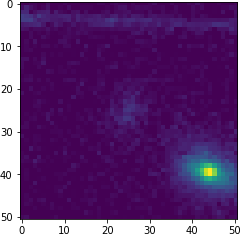}
\includegraphics[width=0.137\linewidth]{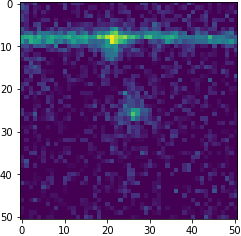}
\includegraphics[width=0.137\linewidth]{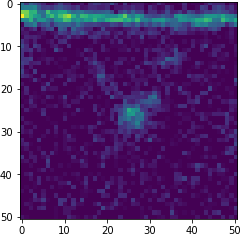}
\includegraphics[width=0.137\linewidth]{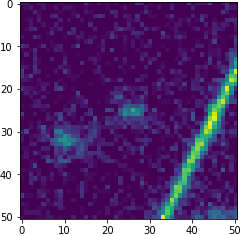}
\includegraphics[width=0.137\linewidth]{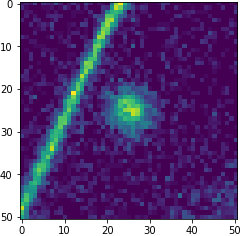}
\includegraphics[width=0.137\linewidth]{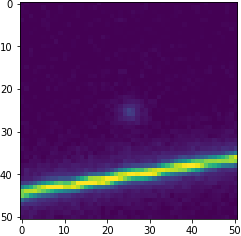}
\includegraphics[width=0.137\linewidth]{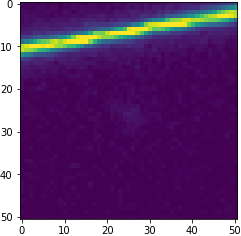}
\includegraphics[width=0.137\linewidth]{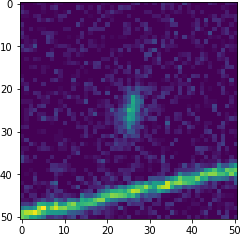}
\includegraphics[width=0.137\linewidth]{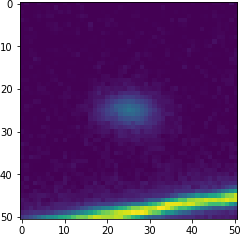}
\includegraphics[width=0.137\linewidth]{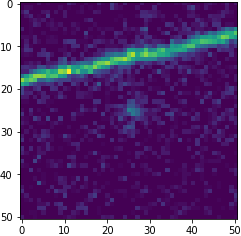}
\includegraphics[width=0.137\linewidth]{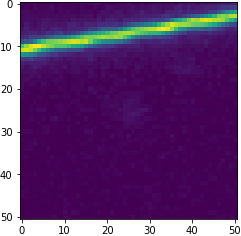}
\includegraphics[width=0.137\linewidth]{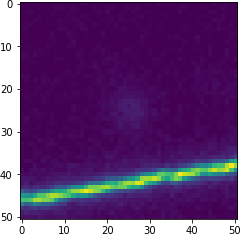}
\includegraphics[width=0.137\linewidth]{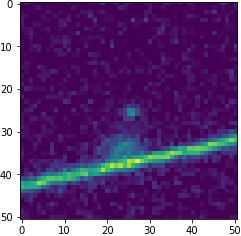}
\includegraphics[width=0.137\linewidth]{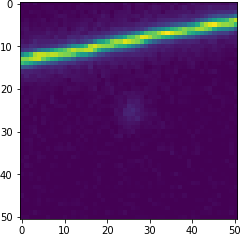}
\includegraphics[width=0.137\linewidth]{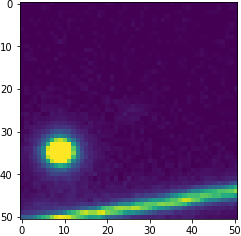}
\includegraphics[width=0.137\linewidth]{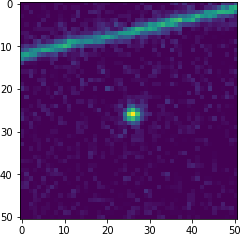}
\includegraphics[width=0.137\linewidth]{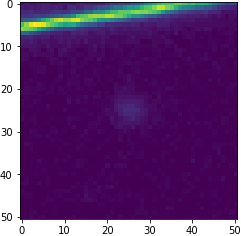}
\includegraphics[width=0.137\linewidth]{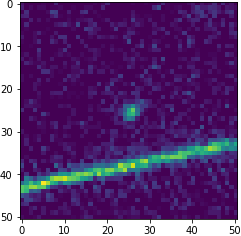}
\includegraphics[width=0.137\linewidth]{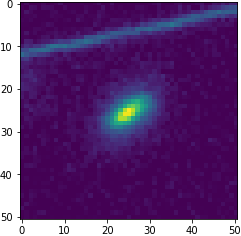}
\includegraphics[width=0.137\linewidth]{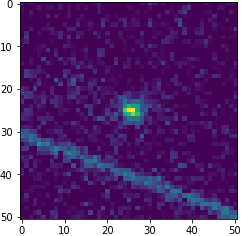}
\includegraphics[width=0.137\linewidth]{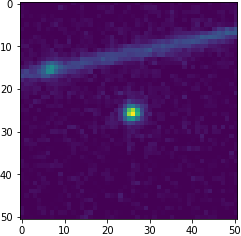}
\includegraphics[width=0.137\linewidth]{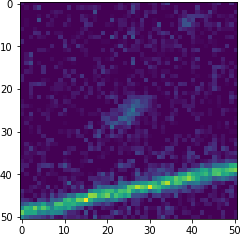}
\includegraphics[width=0.137\linewidth]{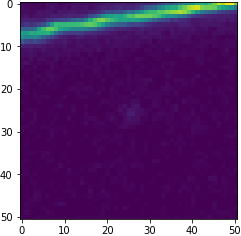}
\includegraphics[width=0.137\linewidth]{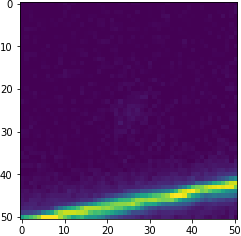}
\includegraphics[width=0.137\linewidth]{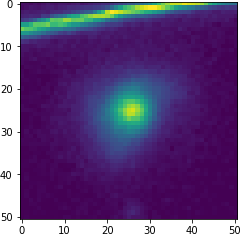}
\includegraphics[width=0.137\linewidth]{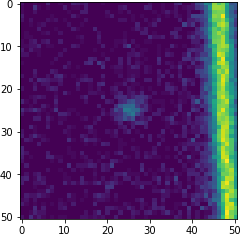}
\includegraphics[width=0.137\linewidth]{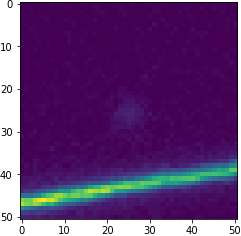}
\includegraphics[width=0.137\linewidth]{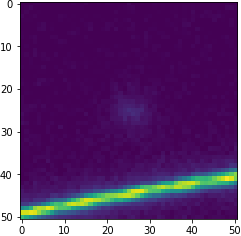}
\includegraphics[width=0.137\linewidth]{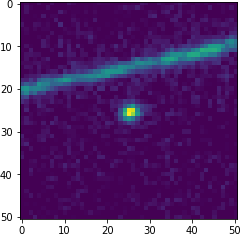}
\includegraphics[width=0.137\linewidth]{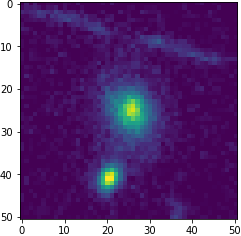}
\includegraphics[width=0.137\linewidth]{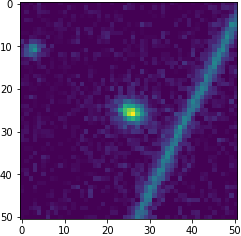}
\includegraphics[width=0.137\linewidth]{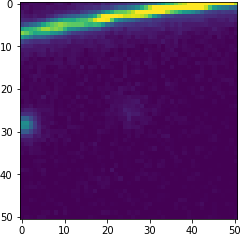}
\includegraphics[width=0.137\linewidth]{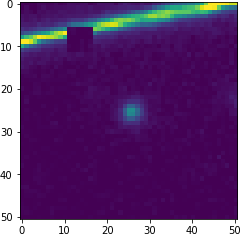}
\includegraphics[width=0.137\linewidth]{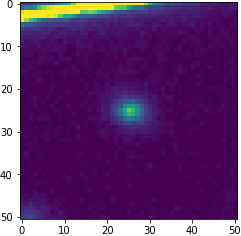}
\includegraphics[width=0.137\linewidth]{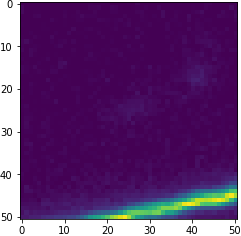}
\includegraphics[width=0.137\linewidth]{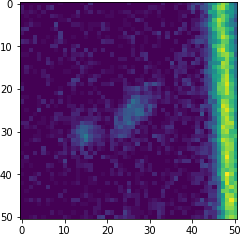}
\includegraphics[width=0.137\linewidth]{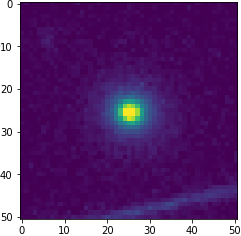}
\includegraphics[width=0.137\linewidth]{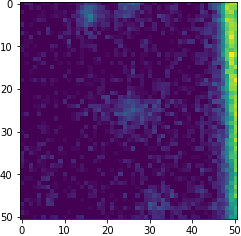}
\includegraphics[width=0.137\linewidth]{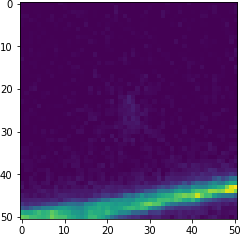}
\includegraphics[width=0.137\linewidth]{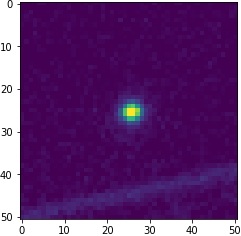}
\includegraphics[width=0.137\linewidth]{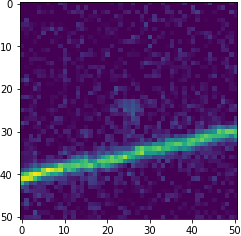}
\includegraphics[width=0.137\linewidth]{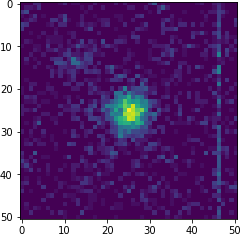}
\includegraphics[width=0.137\linewidth]{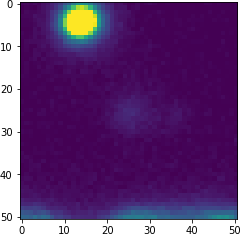}
\includegraphics[width=0.137\linewidth]{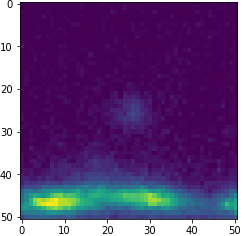}
\includegraphics[width=0.137\linewidth]{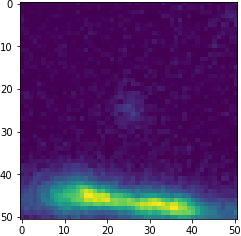}
\includegraphics[width=0.137\linewidth]{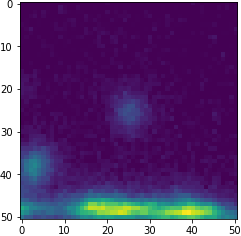}
\includegraphics[width=0.137\linewidth]{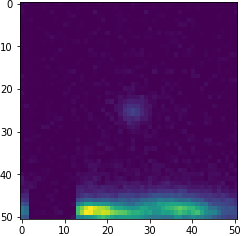}
\includegraphics[width=0.137\linewidth]{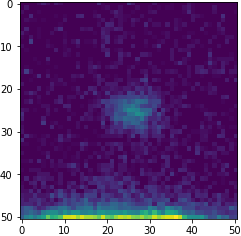}
\includegraphics[width=0.137\linewidth]{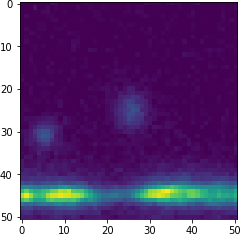}
\includegraphics[width=0.137\linewidth]{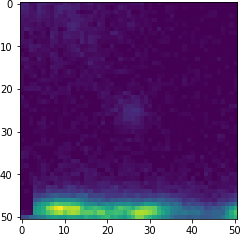}
\includegraphics[width=0.137\linewidth]{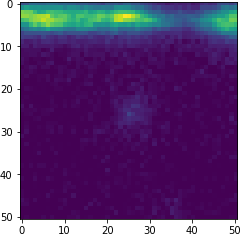}
\includegraphics[width=0.137\linewidth]{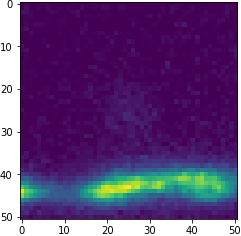}
\includegraphics[width=0.137\linewidth]{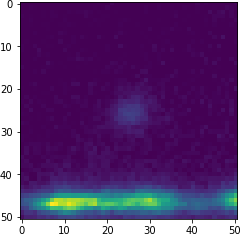}
\caption{continued}

    \label{fig:track2}
\end{figure*}

\begin{figure*}[!hp]
\ContinuedFloat
    \centering
\includegraphics[width=0.137\linewidth]{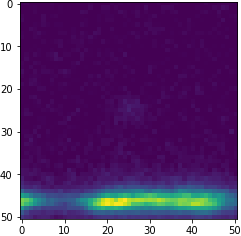} 
\includegraphics[width=0.137\linewidth]{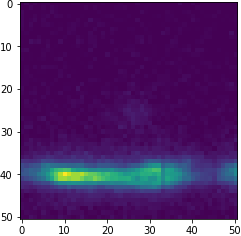} 
\includegraphics[width=0.137\linewidth]{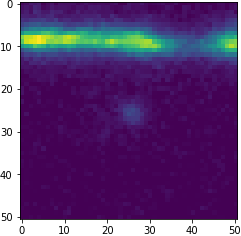}
\includegraphics[width=0.137\linewidth]{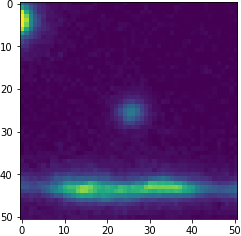}
\includegraphics[width=0.137\linewidth]{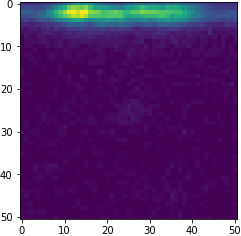}
\includegraphics[width=0.137\linewidth]{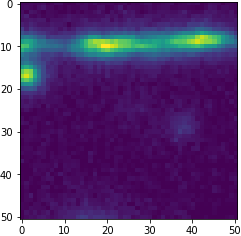}
\includegraphics[width=0.137\linewidth]{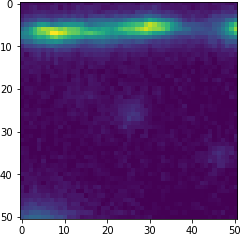}
\includegraphics[width=0.137\linewidth]{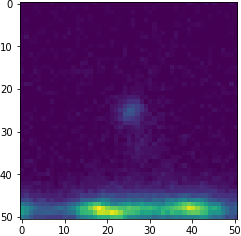}
\includegraphics[width=0.137\linewidth]{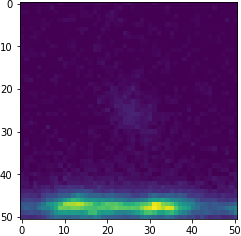}
\includegraphics[width=0.137\linewidth]{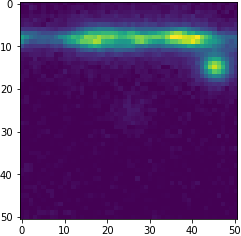}
\includegraphics[width=0.137\linewidth]{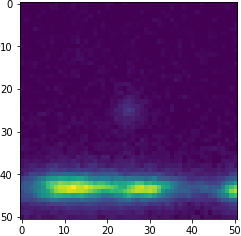}
\includegraphics[width=0.137\linewidth]{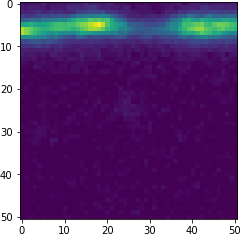}
\includegraphics[width=0.137\linewidth]{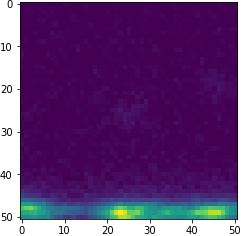}
\includegraphics[width=0.137\linewidth]{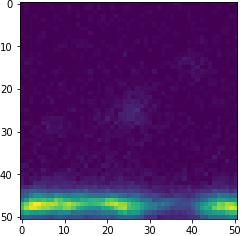}
\includegraphics[width=0.137\linewidth]{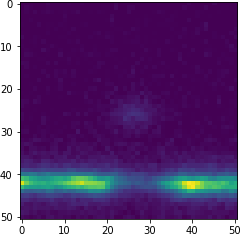}
\includegraphics[width=0.137\linewidth]{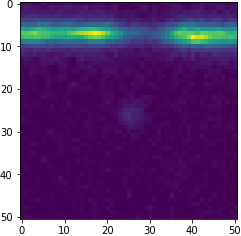}
\includegraphics[width=0.137\linewidth]{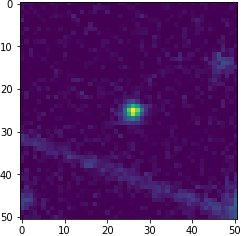}
\includegraphics[width=0.137\linewidth]{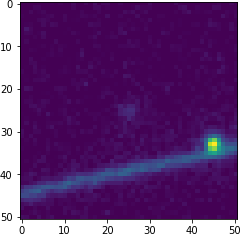}
\includegraphics[width=0.137\linewidth]{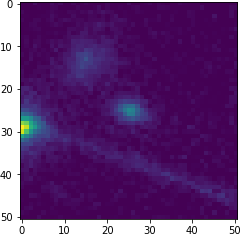}
\includegraphics[width=0.137\linewidth]{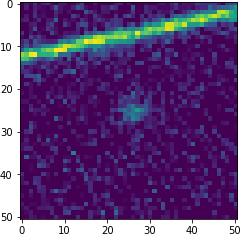}
\includegraphics[width=0.137\linewidth]{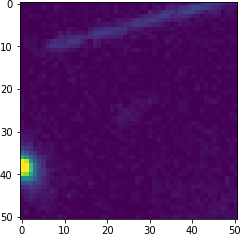}
\includegraphics[width=0.137\linewidth]{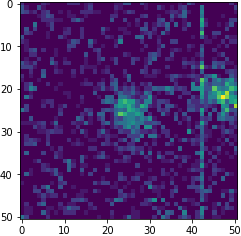}
\includegraphics[width=0.137\linewidth]{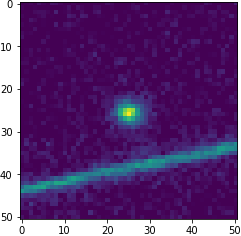}
\includegraphics[width=0.137\linewidth]{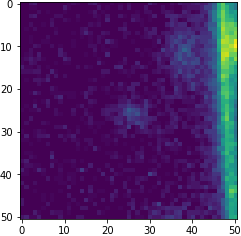}
\includegraphics[width=0.137\linewidth]{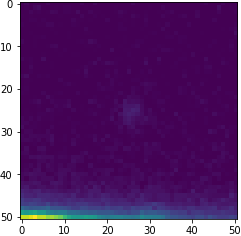}
\includegraphics[width=0.137\linewidth]{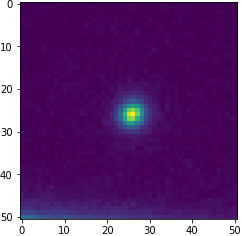}
\includegraphics[width=0.137\linewidth]{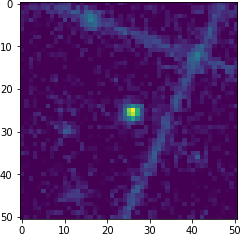}
\includegraphics[width=0.137\linewidth]{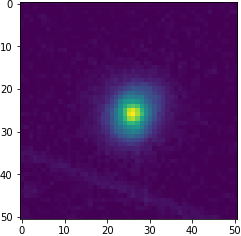}
\includegraphics[width=0.137\linewidth]{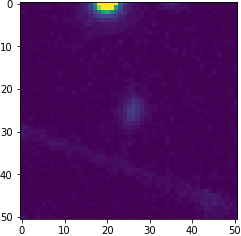}
\includegraphics[width=0.137\linewidth]{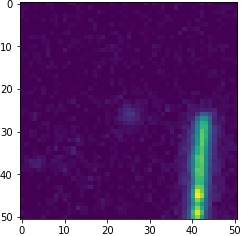}
\includegraphics[width=0.137\linewidth]{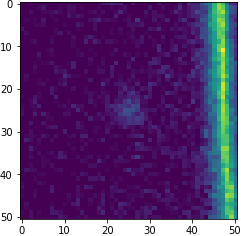}
\includegraphics[width=0.137\linewidth]{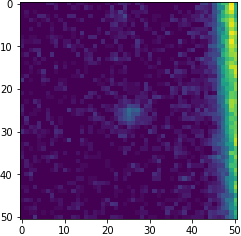}
\includegraphics[width=0.137\linewidth]{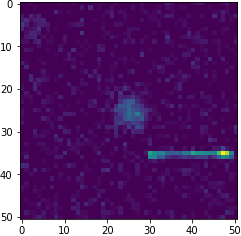}
\includegraphics[width=0.137\linewidth]{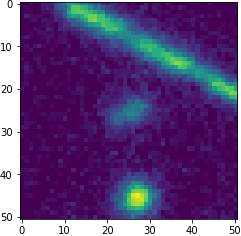}
\includegraphics[width=0.137\linewidth]{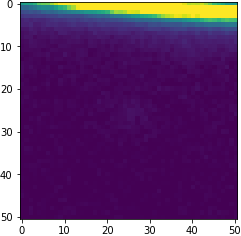}
\includegraphics[width=0.137\linewidth]{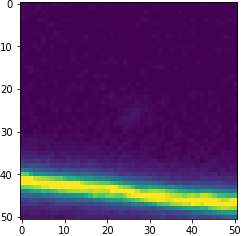}
\includegraphics[width=0.137\linewidth]{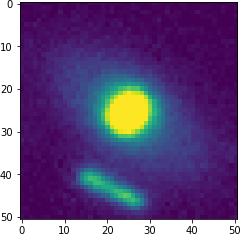}
\includegraphics[width=0.137\linewidth]{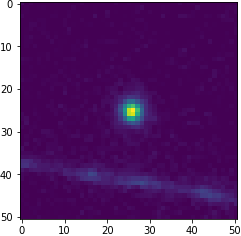}
\includegraphics[width=0.137\linewidth]{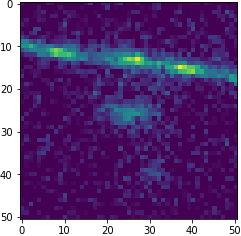}
\includegraphics[width=0.137\linewidth]{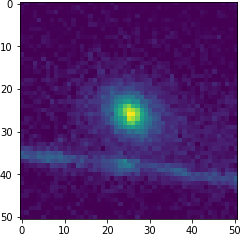}
\includegraphics[width=0.137\linewidth]{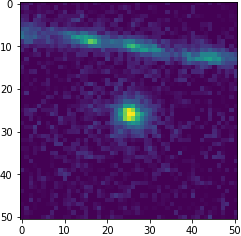}
\includegraphics[width=0.137\linewidth]{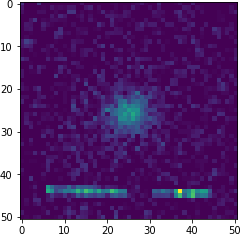}
\includegraphics[width=0.137\linewidth]{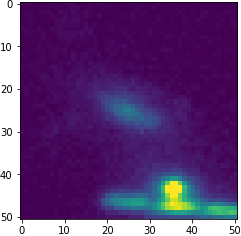}
\includegraphics[width=0.137\linewidth]{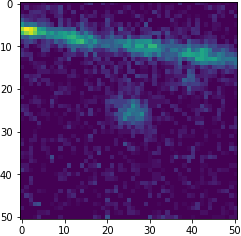}
\includegraphics[width=0.137\linewidth]{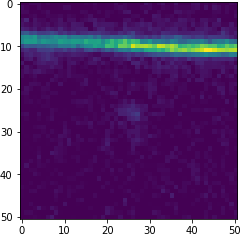}
\includegraphics[width=0.137\linewidth]{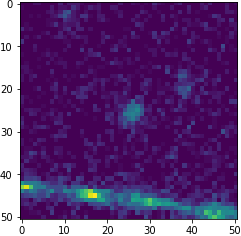}
\includegraphics[width=0.137\linewidth]{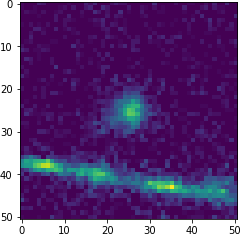}
\includegraphics[width=0.137\linewidth]{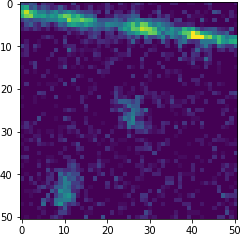}
\includegraphics[width=0.137\linewidth]{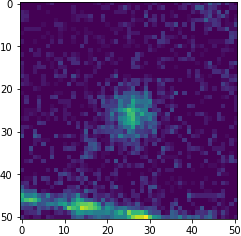}
\includegraphics[width=0.137\linewidth]{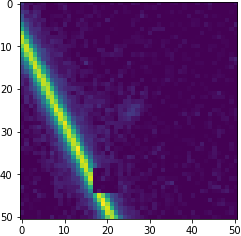}
\includegraphics[width=0.137\linewidth]{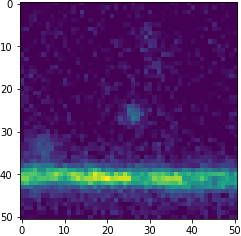}
\includegraphics[width=0.137\linewidth]{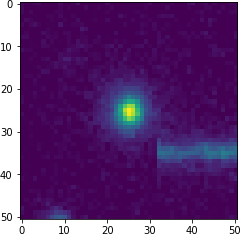}
\includegraphics[width=0.137\linewidth]{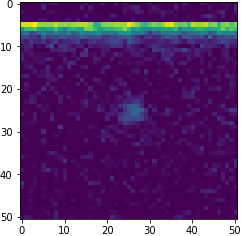}
\includegraphics[width=0.137\linewidth]{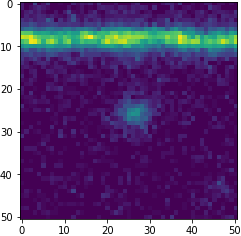}
\includegraphics[width=0.137\linewidth]{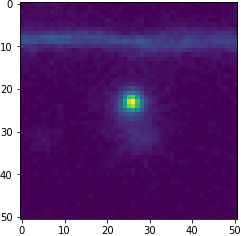}
\includegraphics[width=0.137\linewidth]{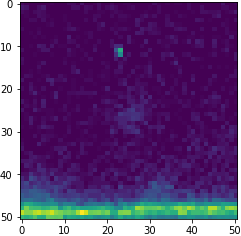}
\caption{continued}

    \label{fig:track3}
\end{figure*}

\begin{figure*}[!hp]
\ContinuedFloat
    \centering
\includegraphics[width=0.138\linewidth]{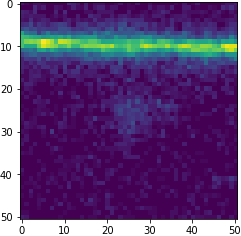}
\includegraphics[width=0.138\linewidth]{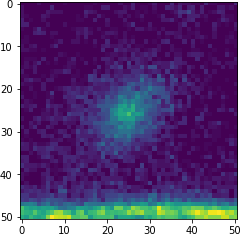}
\includegraphics[width=0.138\linewidth]{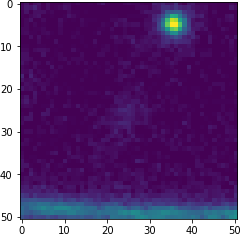}
\includegraphics[width=0.138\linewidth]{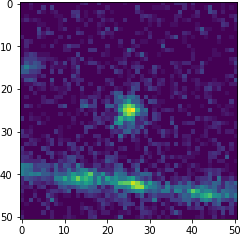}
\includegraphics[width=0.138\linewidth]{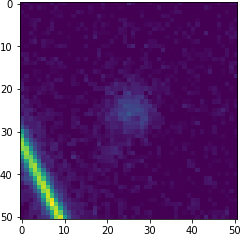}
\includegraphics[width=0.138\linewidth]{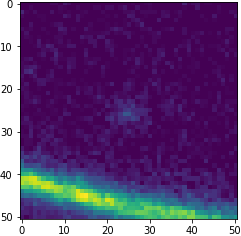}
\includegraphics[width=0.138\linewidth]{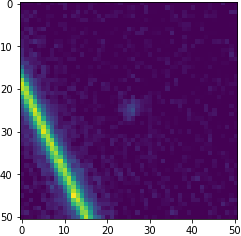}
\includegraphics[width=0.138\linewidth]{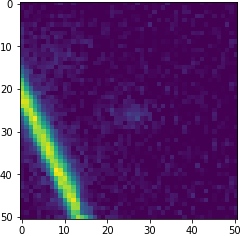}
\includegraphics[width=0.138\linewidth]{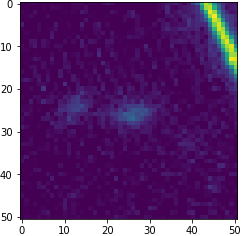}
\includegraphics[width=0.138\linewidth]{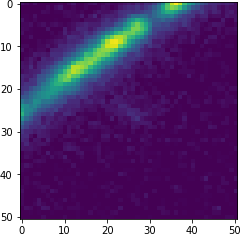}
\includegraphics[width=0.138\linewidth]{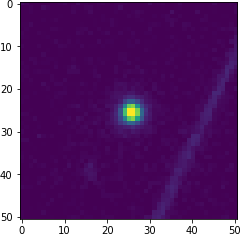}
\includegraphics[width=0.138\linewidth]{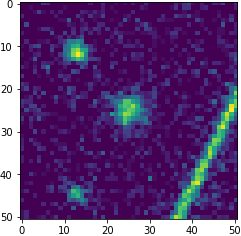}
\includegraphics[width=0.138\linewidth]{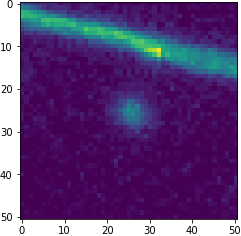}
\includegraphics[width=0.138\linewidth]{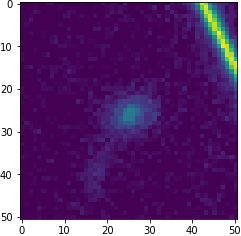}
\includegraphics[width=0.138\linewidth]{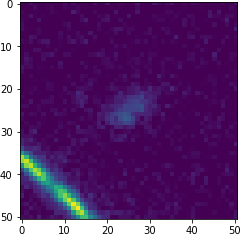}
\includegraphics[width=0.138\linewidth]{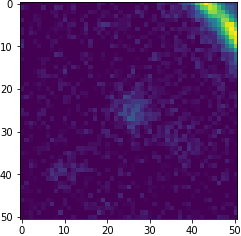}
\includegraphics[width=0.138\linewidth]{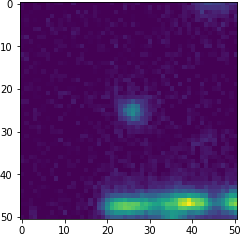}
\caption{continued}

    \label{fig:track4}
\end{figure*}

\begin{figure*}[!hp]
    \centering
\includegraphics[width=.138\linewidth]{problematic/10523_46.png}
\includegraphics[width=.138\linewidth]{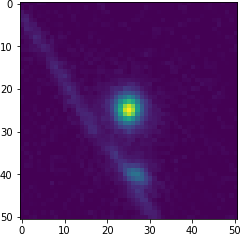}
\includegraphics[width=.138\linewidth]{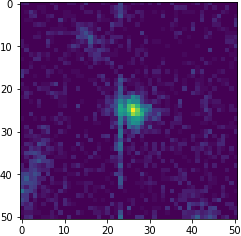}
\includegraphics[width=.138\linewidth]{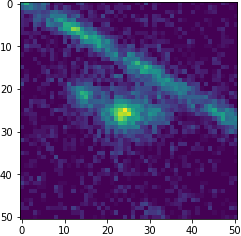}
\includegraphics[width=.138\linewidth]{problematic/13077_47.png}
\includegraphics[width=.138\linewidth]{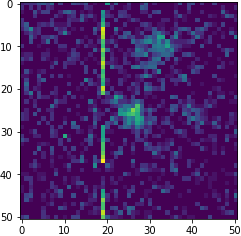}
\includegraphics[width=.138\linewidth]{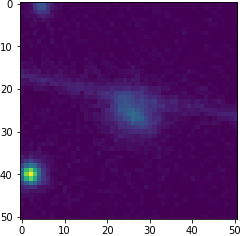} 
\includegraphics[width=.138\linewidth]{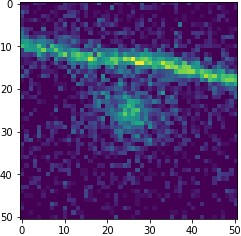}
\includegraphics[width=.138\linewidth]{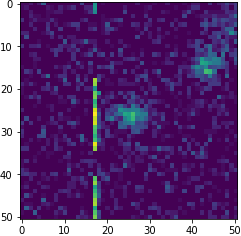}
\includegraphics[width=.138\linewidth]{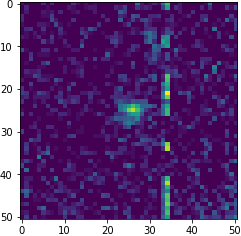}
\includegraphics[width=.138\linewidth]{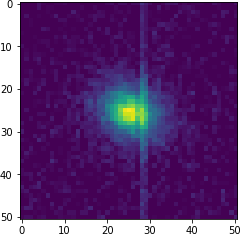}
\includegraphics[width=.138\linewidth]{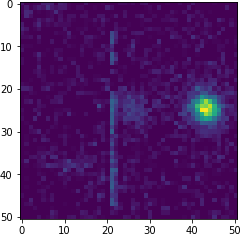}
\includegraphics[width=.138\linewidth]{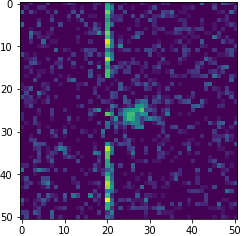}
\includegraphics[width=.138\linewidth]{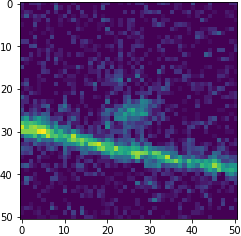}
\includegraphics[width=.138\linewidth]{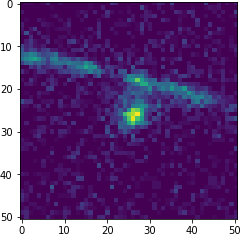}
\includegraphics[width=.138\linewidth]{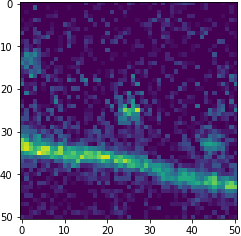}
\includegraphics[width=.138\linewidth]{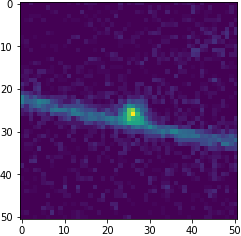} 
\includegraphics[width=.138\linewidth]{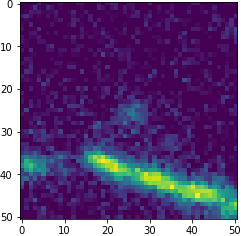}
\includegraphics[width=.138\linewidth]{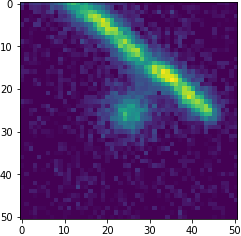}
\includegraphics[width=.138\linewidth]{problematic/16734_17.png}
\includegraphics[width=.138\linewidth]{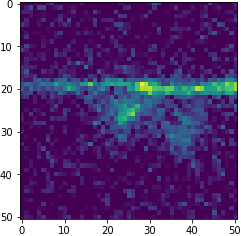}
\includegraphics[width=.138\linewidth]{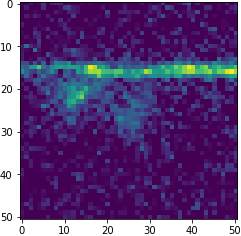}
\includegraphics[width=.138\linewidth]{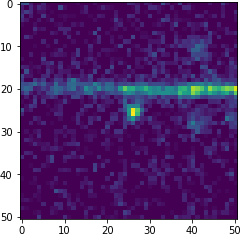} 
\includegraphics[width=.138\linewidth]{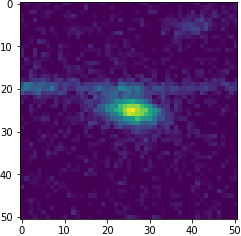} 
\includegraphics[width=.138\linewidth]{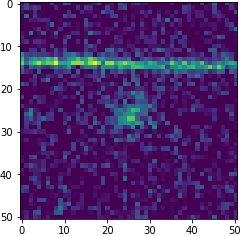}
\includegraphics[width=.138\linewidth]{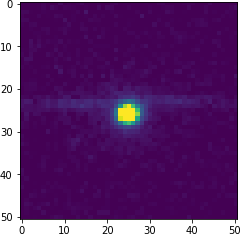} 
\includegraphics[width=.138\linewidth]{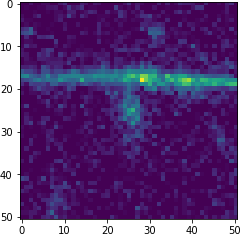}
\includegraphics[width=.138\linewidth]{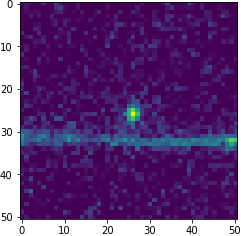} 
\includegraphics[width=.138\linewidth]{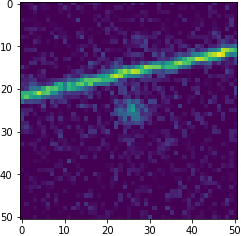} 
\includegraphics[width=.138\linewidth]{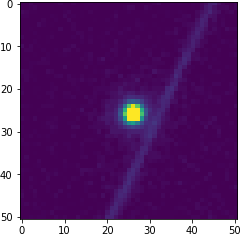} 
\includegraphics[width=.138\linewidth]{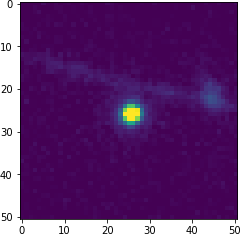}
\includegraphics[width=.138\linewidth]{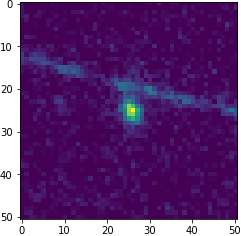}
\includegraphics[width=.138\linewidth]{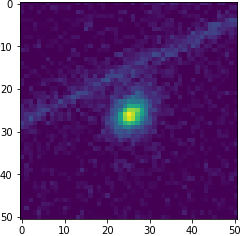}
\includegraphics[width=.138\linewidth]{problematic/19347_05.png}
\includegraphics[width=.138\linewidth]{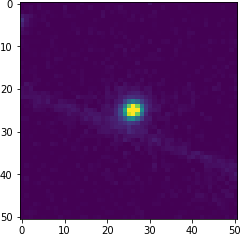}
\includegraphics[width=.138\linewidth]{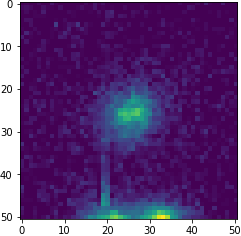}
\includegraphics[width=.138\linewidth]{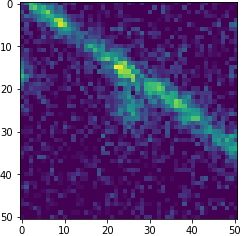}
\includegraphics[width=.138\linewidth]{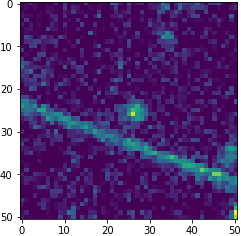}
\includegraphics[width=.138\linewidth]{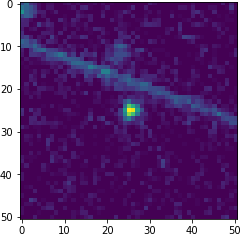}
\includegraphics[width=.138\linewidth]{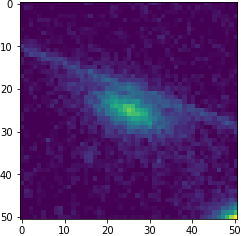}
\includegraphics[width=.138\linewidth]{problematic/22797_44.png}
\includegraphics[width=.138\linewidth]{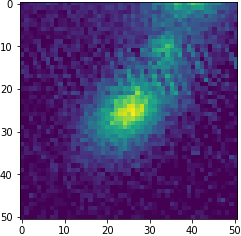}
\includegraphics[width=.138\linewidth]{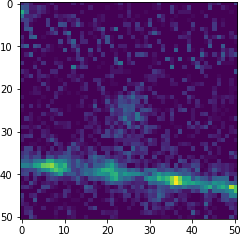}
\includegraphics[width=.138\linewidth]{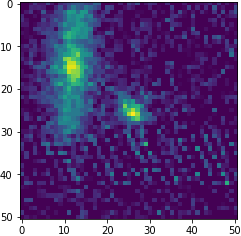}
\includegraphics[width=.138\linewidth]{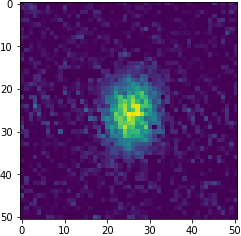}
\includegraphics[width=.138\linewidth]{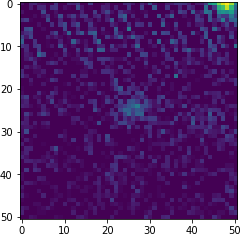}
\includegraphics[width=.138\linewidth]{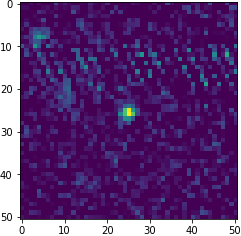}
\includegraphics[width=.138\linewidth]{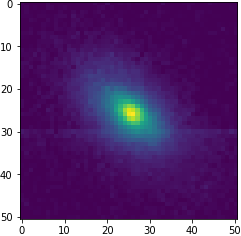}
\includegraphics[width=.138\linewidth]{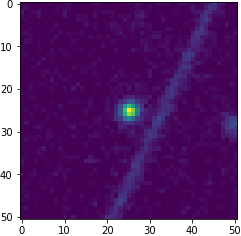}
\includegraphics[width=.138\linewidth]{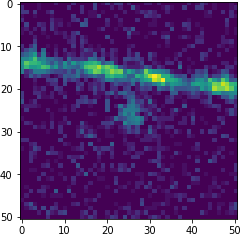}
\includegraphics[width=.138\linewidth]{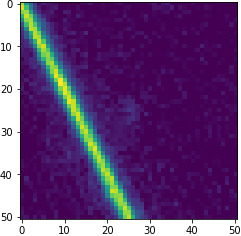} 
\includegraphics[width=.138\linewidth]{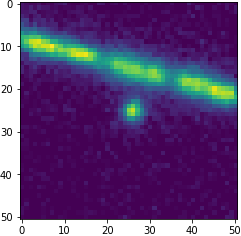} 
\caption{Problematic objects. In most of the cases, if a track appears, it is closer to the object with respect to the aperture photometry, so it slightly affects the flux measurement.}
    \label{fig:problematic-app}
\end{figure*}

\begin{figure*}[!hp]
    \centering
\includegraphics[width=.138\linewidth]{lc/lc10523.png}
\includegraphics[width=.138\linewidth]{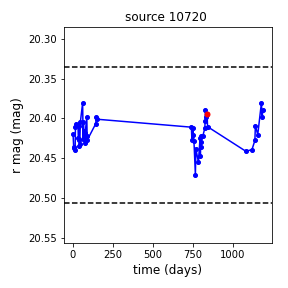}
\includegraphics[width=.138\linewidth]{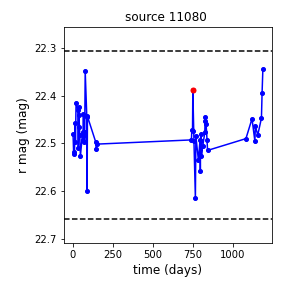}
\includegraphics[width=.138\linewidth]{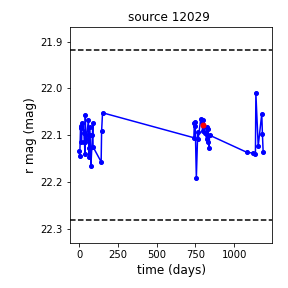}
\includegraphics[width=.138\linewidth]{lc/lc13077.png}
\includegraphics[width=.138\linewidth]{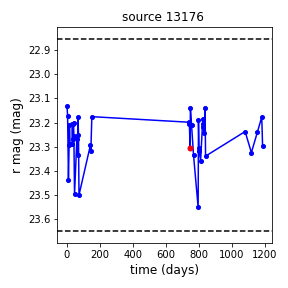}
\includegraphics[width=.138\linewidth]{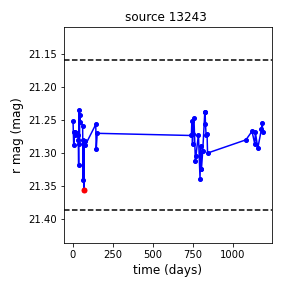} 
\includegraphics[width=.138\linewidth]{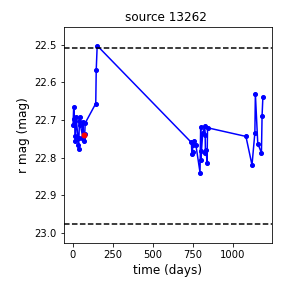}
\includegraphics[width=.138\linewidth]{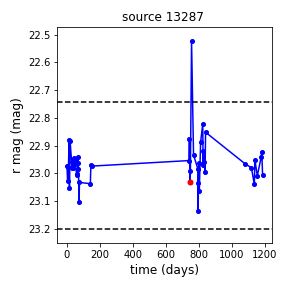}
\includegraphics[width=.138\linewidth]{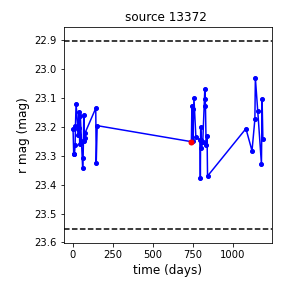}
\includegraphics[width=.138\linewidth]{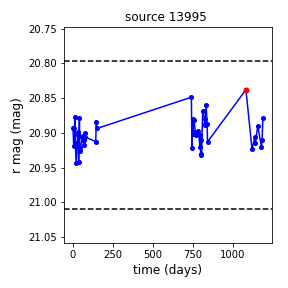}
\includegraphics[width=.138\linewidth]{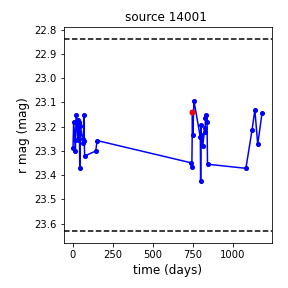}
\includegraphics[width=.138\linewidth]{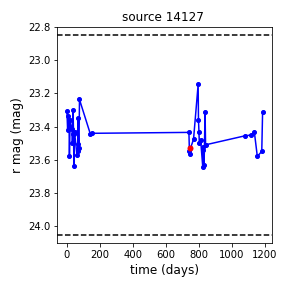}
\includegraphics[width=.138\linewidth]{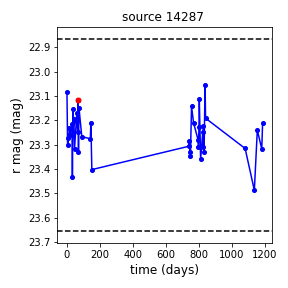}
\includegraphics[width=.138\linewidth]{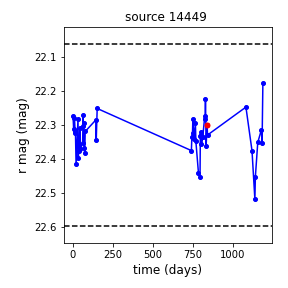}
\includegraphics[width=.138\linewidth]{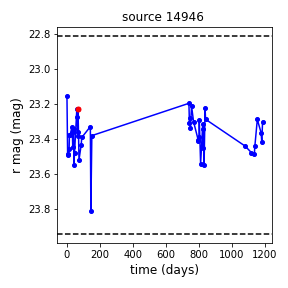}
\includegraphics[width=.138\linewidth]{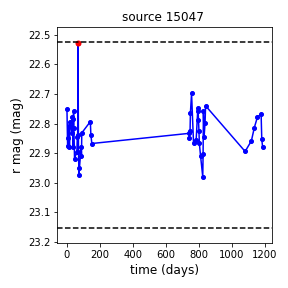} 
\includegraphics[width=.138\linewidth]{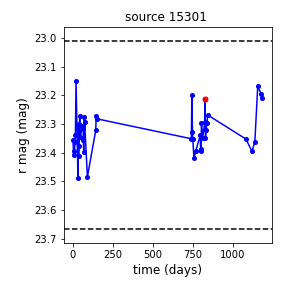}
\includegraphics[width=.138\linewidth]{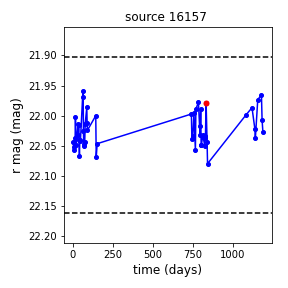}
\includegraphics[width=.138\linewidth]{lc/lc16734.png}
\includegraphics[width=.138\linewidth]{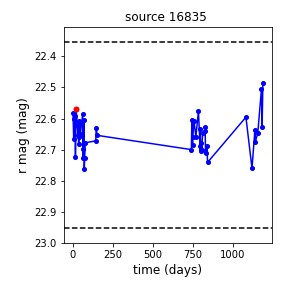}
\includegraphics[width=.138\linewidth]{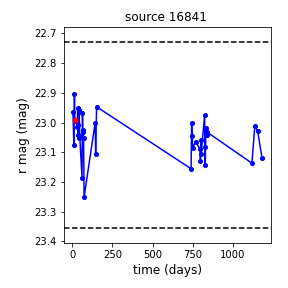}
\includegraphics[width=.138\linewidth]{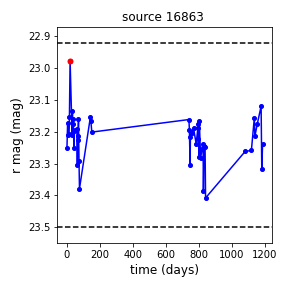} 
\includegraphics[width=.138\linewidth]{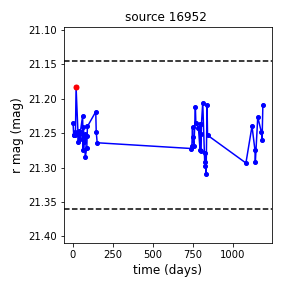} 
\includegraphics[width=.138\linewidth]{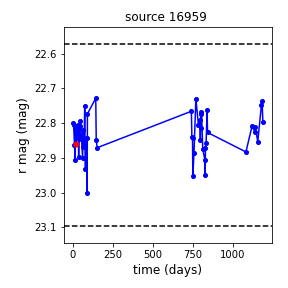}
\includegraphics[width=.138\linewidth]{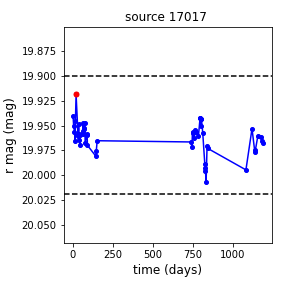} 
\includegraphics[width=.138\linewidth]{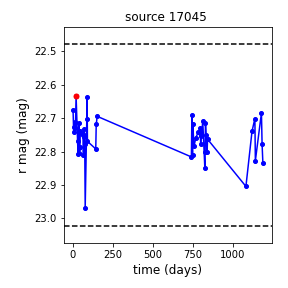}
\includegraphics[width=.138\linewidth]{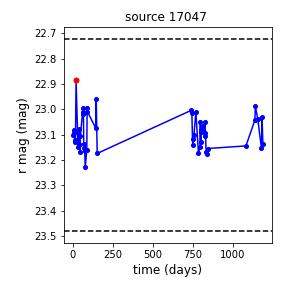} 
\includegraphics[width=.138\linewidth]{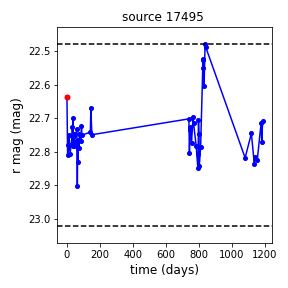}
\includegraphics[width=.138\linewidth]{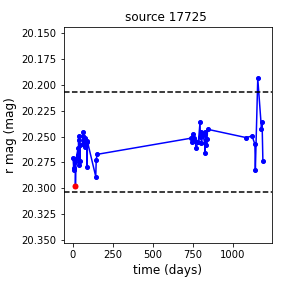} 
\includegraphics[width=.138\linewidth]{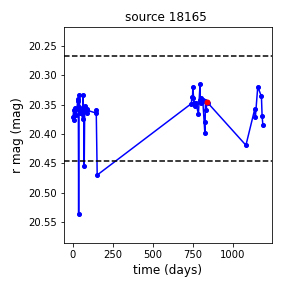}
\includegraphics[width=.138\linewidth]{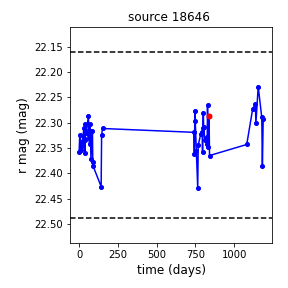}
\includegraphics[width=.138\linewidth]{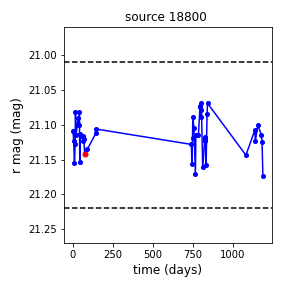}
\includegraphics[width=.138\linewidth]{lc/lc19347.png}
\includegraphics[width=.138\linewidth]{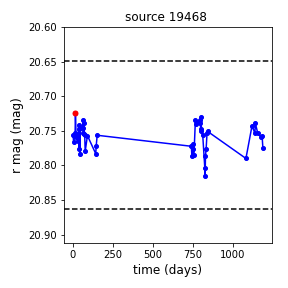}
\includegraphics[width=.138\linewidth]{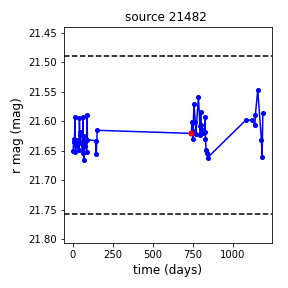}
\includegraphics[width=.138\linewidth]{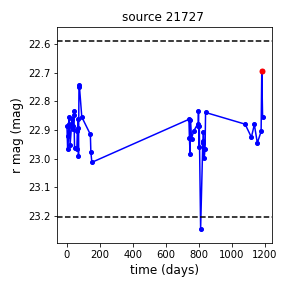}
\includegraphics[width=.138\linewidth]{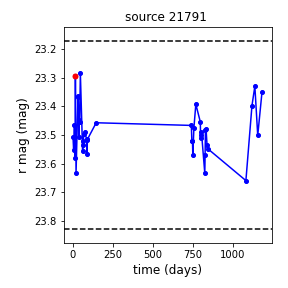}
\includegraphics[width=.138\linewidth]{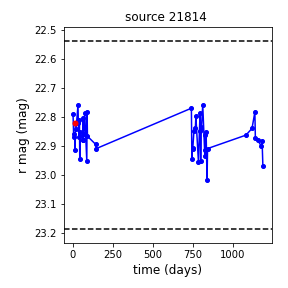}
\includegraphics[width=.138\linewidth]{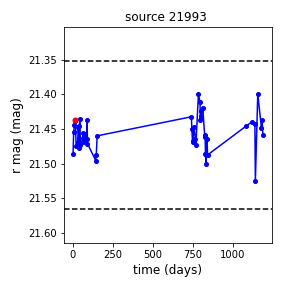}
\includegraphics[width=.138\linewidth]{lc/lc22797.png}
\includegraphics[width=.138\linewidth]{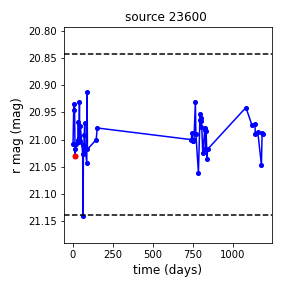}
\includegraphics[width=.138\linewidth]{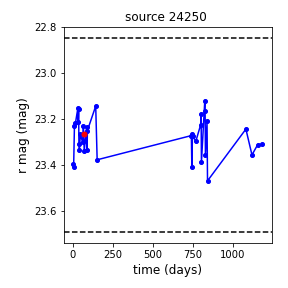}
\includegraphics[width=.138\linewidth]{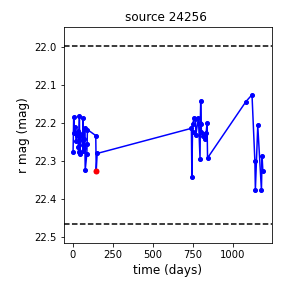}
\includegraphics[width=.138\linewidth]{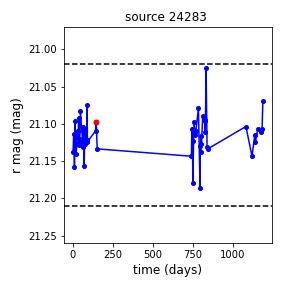}
\includegraphics[width=.138\linewidth]{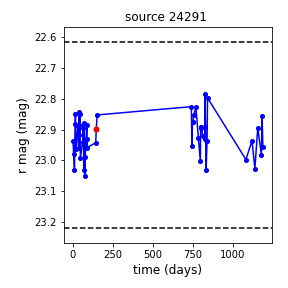}
\includegraphics[width=.138\linewidth]{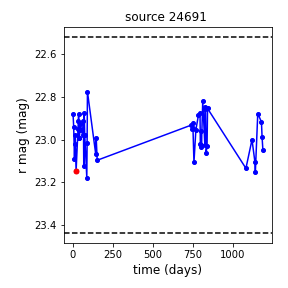}
\includegraphics[width=.138\linewidth]{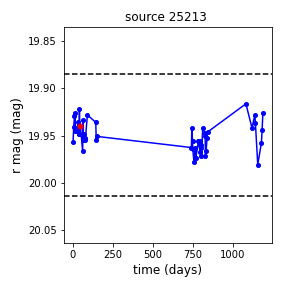}
\includegraphics[width=.138\linewidth]{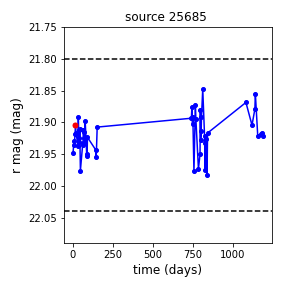}
\includegraphics[width=.138\linewidth]{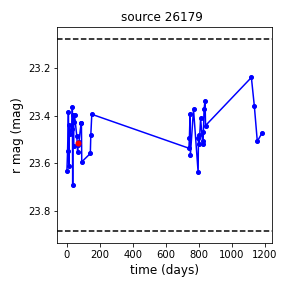}
\includegraphics[width=.138\linewidth]{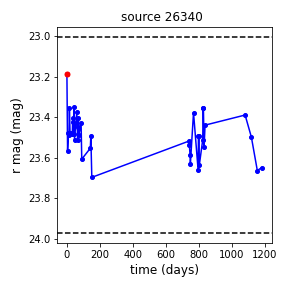} 
\includegraphics[width=.138\linewidth]{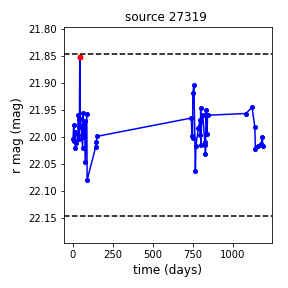} 
\caption{Light curves for the 52 problematic objects shown in Fig.~\ref{fig:problematic-app}. The flagged epoch is marked in red. The horizontal lines represent the threshold of the sigma clip. }
    \label{fig:problematic-lc-app}
\end{figure*}

\begin{figure*}[!hp]
    \centering
\includegraphics[width=0.138\linewidth]{bohBefAft/12104_34.png}
\includegraphics[width=0.138\linewidth]{bohBefAft/12104_35.png}
\includegraphics[width=0.138\linewidth]{bohBefAft/12104_36.png}
\hfill
\includegraphics[width=0.138\linewidth]{bohBefAft/12593_49.png}
\includegraphics[width=0.138\linewidth]{bohBefAft/12593_50.png}
\includegraphics[width=0.138\linewidth]{bohBefAft/12593_51.png}
\\
\includegraphics[width=0.138\linewidth]{bohBefAft/13243_30.png}
\includegraphics[width=0.138\linewidth]{bohBefAft/13243_31.png}
\includegraphics[width=0.138\linewidth]{bohBefAft/13243_32.png}
\hfill
\includegraphics[width=0.138\linewidth]{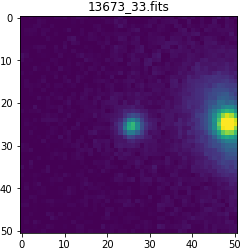}
\includegraphics[width=0.138\linewidth]{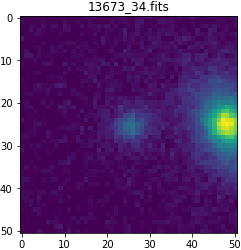}
\includegraphics[width=0.138\linewidth]{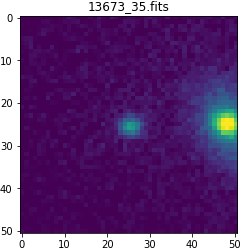}
\\
\includegraphics[width=0.138\linewidth]{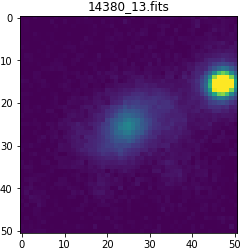}
\includegraphics[width=0.138\linewidth]{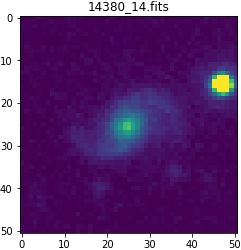}
\includegraphics[width=0.138\linewidth]{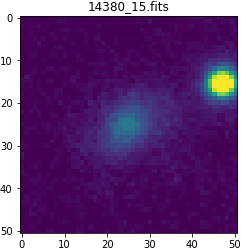}
\hfill
\includegraphics[width=0.138\linewidth]{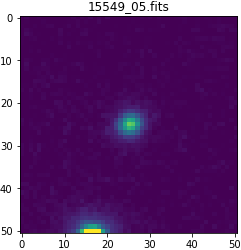}
\includegraphics[width=0.138\linewidth]{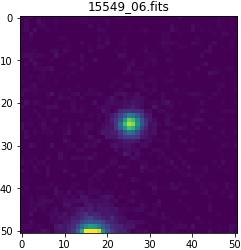}
\includegraphics[width=0.138\linewidth]{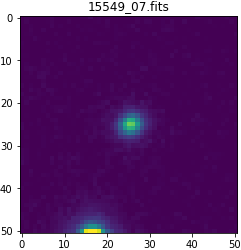}
\\
\includegraphics[width=0.138\linewidth]{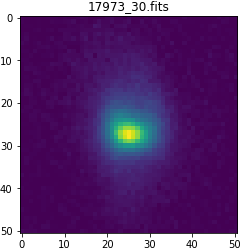}
\includegraphics[width=0.138\linewidth]{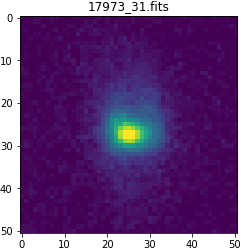}
\includegraphics[width=0.138\linewidth]{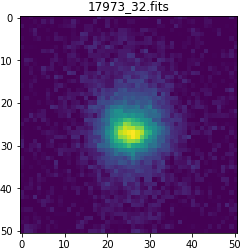}
\hfill
\includegraphics[width=0.138\linewidth]{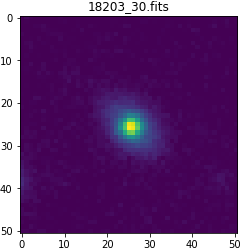}
\includegraphics[width=0.138\linewidth]{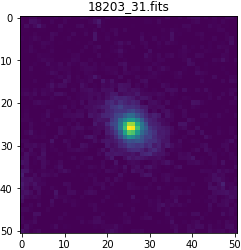}
\includegraphics[width=0.138\linewidth]{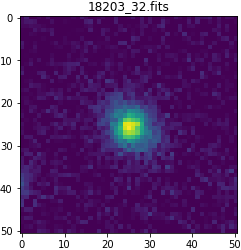}
\\
\includegraphics[width=0.138\linewidth]{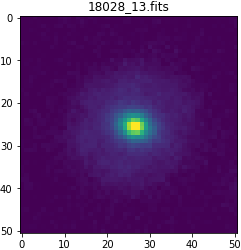}
\includegraphics[width=0.138\linewidth]{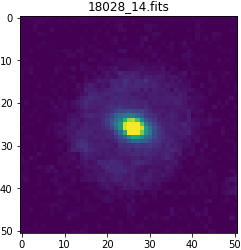}
\includegraphics[width=0.138\linewidth]{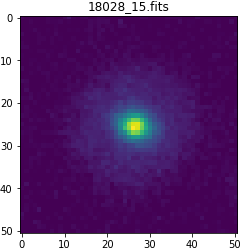}
\hfill
\includegraphics[width=0.138\linewidth]{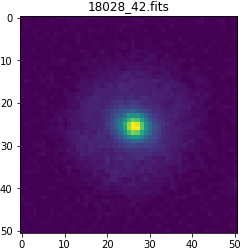}
\includegraphics[width=0.138\linewidth]{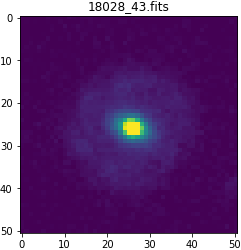}
\includegraphics[width=0.138\linewidth]{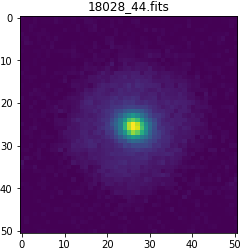}
\\
\includegraphics[width=0.138\linewidth]{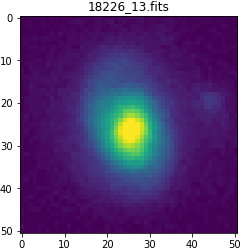}
\includegraphics[width=0.138\linewidth]{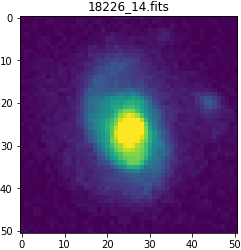}
\includegraphics[width=0.138\linewidth]{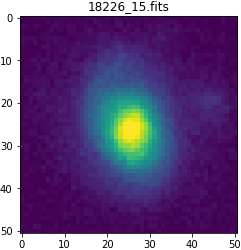}
\hfill
\includegraphics[width=0.138\linewidth]{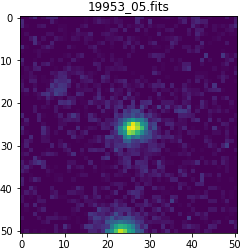}
\includegraphics[width=0.138\linewidth]{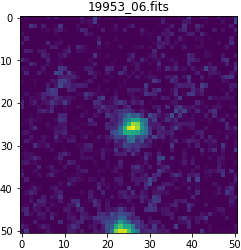}
\includegraphics[width=0.138\linewidth]{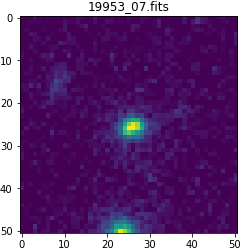}
\\
\includegraphics[width=0.138\linewidth]{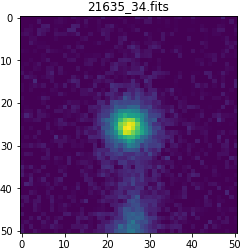}
\includegraphics[width=0.138\linewidth]{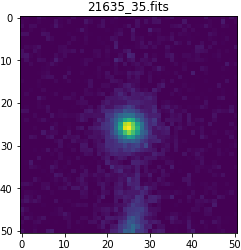}
\includegraphics[width=0.138\linewidth]{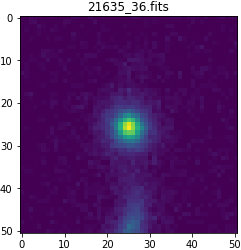}
\hfill
\includegraphics[width=0.138\linewidth]{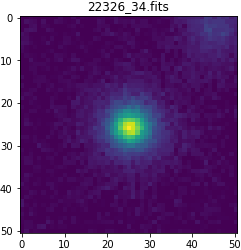}
\includegraphics[width=0.138\linewidth]{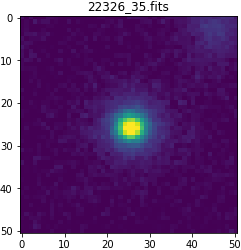}
\includegraphics[width=0.138\linewidth]{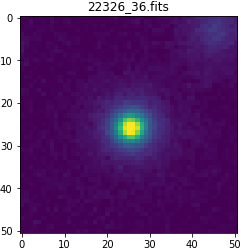}
\\
\caption{Objects flagged as anomalies for which we do not have a proper motivation. Some of the anomalies can still be low S/N objects. For each epoch flagged, we added the one before and the one after for comparison. In two cases (second line, left column and fourth line, right column), epoch 32 is selected as the ``after'' epoch.}
\label{fig:boh1-app}
\end{figure*}

\begin{figure*}[!hp]
    \centering
\includegraphics[width=0.138\linewidth]{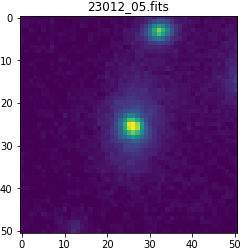}
\includegraphics[width=0.138\linewidth]{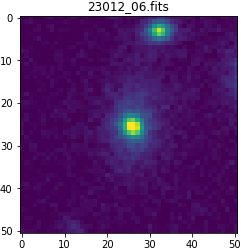}
\includegraphics[width=0.138\linewidth]{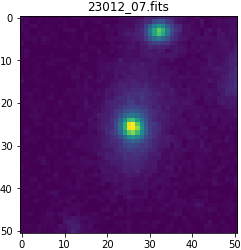}
\hfill
\includegraphics[width=0.138\linewidth]{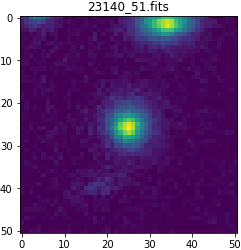}
\includegraphics[width=0.138\linewidth]{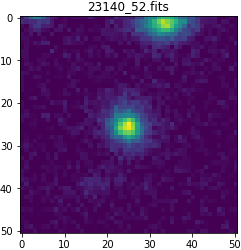}
\includegraphics[width=0.138\linewidth]{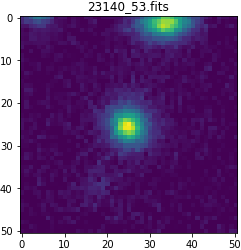}
\\
\includegraphics[width=0.138\linewidth]{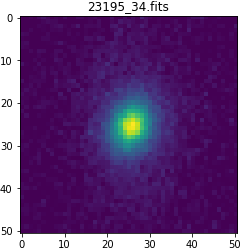}
\includegraphics[width=0.138\linewidth]{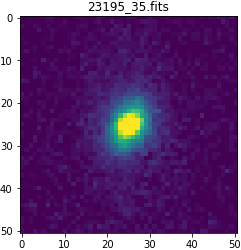}
\includegraphics[width=0.138\linewidth]{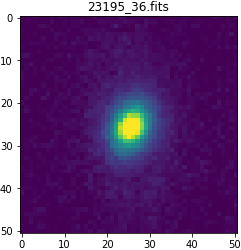}
\hfill
\includegraphics[width=0.138\linewidth]{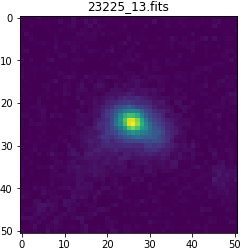}
\includegraphics[width=0.138\linewidth]{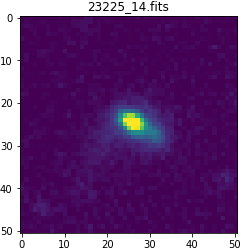}
\includegraphics[width=0.138\linewidth]{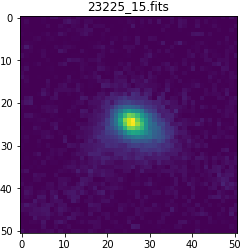}
\\
\includegraphics[width=0.138\linewidth]{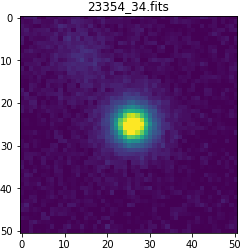}
\includegraphics[width=0.138\linewidth]{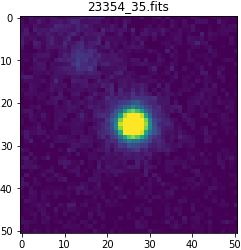}
\includegraphics[width=0.138\linewidth]{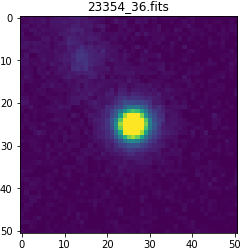}
\hfill
\includegraphics[width=0.138\linewidth]{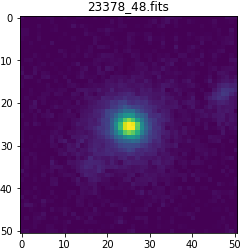}
\includegraphics[width=0.138\linewidth]{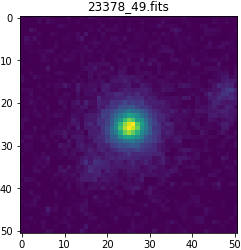}
\includegraphics[width=0.138\linewidth]{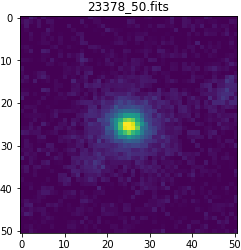}
\\
\includegraphics[width=0.138\linewidth]{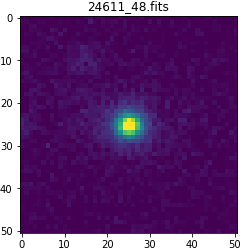}
\includegraphics[width=0.138\linewidth]{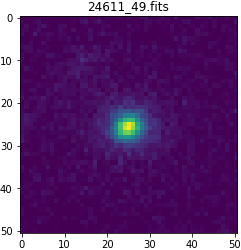}
\includegraphics[width=0.138\linewidth]{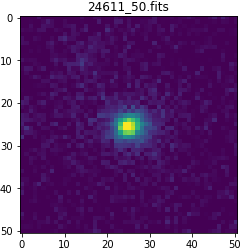}
\hfill
\includegraphics[width=0.138\linewidth]{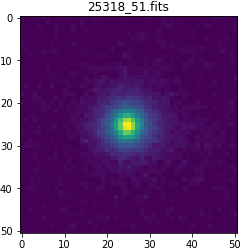}
\includegraphics[width=0.138\linewidth]{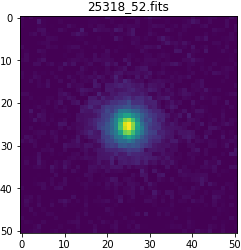}
\includegraphics[width=0.138\linewidth]{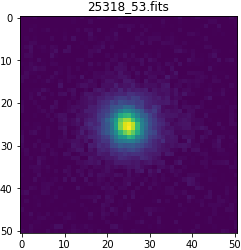}
\\
\includegraphics[width=0.138\linewidth]{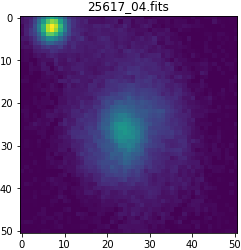}
\includegraphics[width=0.138\linewidth]{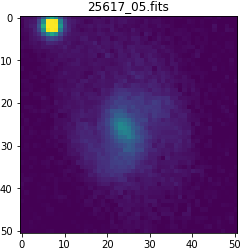}
\includegraphics[width=0.138\linewidth]{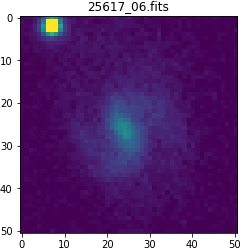}
\hfill
\includegraphics[width=0.138\linewidth]{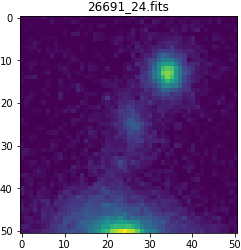}
\includegraphics[width=0.138\linewidth]{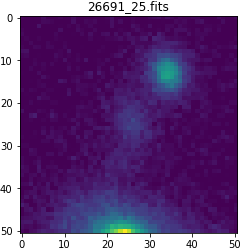}
\includegraphics[width=0.138\linewidth]{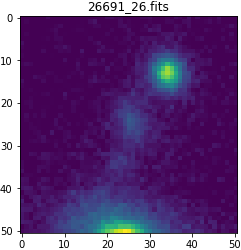}
\\

\includegraphics[width=0.138\linewidth]{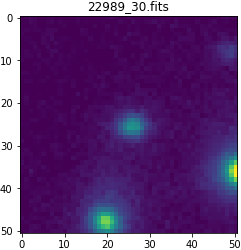}
\includegraphics[width=0.138\linewidth]{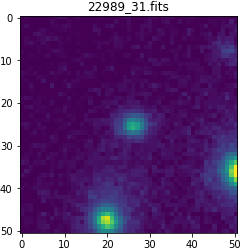}
\includegraphics[width=0.138\linewidth]{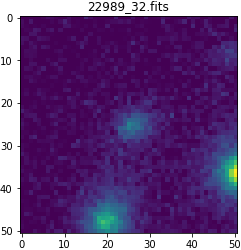}
\hfill
\includegraphics[width=0.138\linewidth]{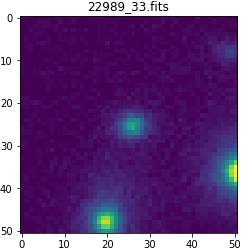}
\includegraphics[width=0.138\linewidth]{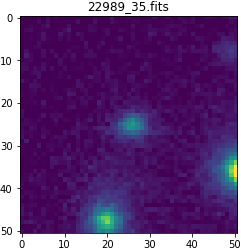}
\includegraphics[width=0.138\linewidth]{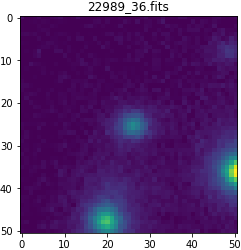}\\
\includegraphics[width=0.138\linewidth]{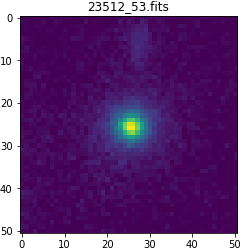}
\includegraphics[width=0.138\linewidth]{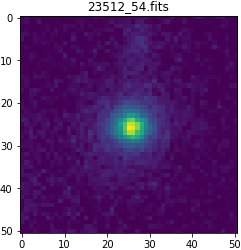}
\caption{Objects flagged as anomalies for which we do not have a proper motivation. Some of the anomalies can still be low S/N objects. For each epoch flagged, we added the one before and the one after for comparison. In the last two lines, six epochs appear since the method selected more than one epoch (epochs 31 and 35; 34 was removed by the sigma clipping), and this causes a series (from epoch 30 to 36). In the last line, there are only two epochs since the selected one is the last.}
\label{fig:boh2}
\end{figure*}

\end{document}